\documentclass[%
10pt,
onecolumn,
superscriptaddress,
preprintnumbers,
nofootinbib,
 amsmath,amssymb,
 aps,
prd,
longbibliography,
notitlepage
]{revtex4-2}
\usepackage{amsmath,amssymb,bm,color,graphicx}
\usepackage[mathscr]{eucal}
\usepackage{slashed}
\usepackage{cancel}
\usepackage[caption=false]{subfig}
\usepackage{tikz,xcolor}
\usepackage{multirow}
\usepackage{tensor}
\usepackage{mathtools}
\usepackage[colorlinks=true,linkcolor=blue,citecolor=blue,urlcolor=blue]{hyperref}

\definecolor{codegreen}{rgb}{0,0.6,0}
\definecolor{codegray}{rgb}{0.5,0.5,0.5}
\definecolor{codepurple}{rgb}{0.58,0,0.82}
\definecolor{backcolour}{rgb}{0.95,0.95,0.92}

\newcommand{\ed}{\mathcal{E}}

\begin{document}

\preprint{RIKEN-iTHEMS-Report-26, YITP-26-52}

\author{Kohei Fujikura}
\email{kohei.fujikura@yukawa.kyoto-u.ac.jp} 
\affiliation{Yukawa Institute for Theoretical Physics, Kyoto University, Kyoto 606-8502, Japan}

\author{Yoshimasa Hidaka}
\email{yoshimasa.hidaka@yukawa.kyoto-u.ac.jp}

\affiliation{Yukawa Institute for Theoretical Physics, Kyoto University, Kyoto 606-8502, Japan}
\affiliation{RIKEN Center for Interdisciplinary Theoretical and Mathematical Sciences (iTHEMS), RIKEN, Wako 351-0198, Japan}

\title{
Dense \texorpdfstring{$\mathrm{QC_2D_2}$}{QC2D2} with uniform matrix product states
}
\begin{abstract}
We study cold dense single-flavor $\mathrm{SU}(2)$ gauge theory in $(1+1)$ dimensions in the thermodynamic limit using a gauge-invariant variational uniform matrix product state ansatz. This formulation provides a sign-problem-free, first-principles approach to dense QCD. We show that, at finite baryon density, the infrared behavior is consistent with a Tomonaga--Luttinger liquid: the central charge is determined to be $c=1$, and the two-point function of the baryon-number density exhibits spatial modulation with the wavenumber predicted by Tomonaga--Luttinger liquid theory. The Luttinger parameter varies smoothly from $K\simeq 1$ in the dilute-baryon regime to $K\simeq 1/2$ at higher densities, suggesting a quarkyonic crossover. Furthermore, the quark distribution reveals the coexistence of a quark Fermi sea with a baryonic infrared description, thereby realizing the quarkyonic picture from first principles.
\end{abstract}

\maketitle

\def\thefootnote{\arabic{footnote}}
\setcounter{footnote}{0}

\tableofcontents

\section{Introduction}

Revealing the phase structure of cold dense quantum chromodynamics (QCD) is one of the most important topics for understanding the physics of neutron stars.
Various phases of dense QCD, including the quarkyonic matter~\cite{McLerran:2007qj,Kojo:2009ha,Kojo:2011cn} and color superconductivity~\cite{Alford:2007xm} have been proposed so far (See e.g., refs.~\cite{Fukushima:2010bq,Fukushima:2013rx,Fischer:2018sdj,McLerran:2026dio} for reviews of the dense QCD phase diagram).
However, there is no direct evidence from first-principles approaches, except at very large baryon chemical potential, where perturbative QCD becomes applicable.
Although conventional Monte-Carlo simulations are a powerful tool for investigating the non-perturbative nature of QCD, they typically suffer from the infamous sign problem when a $\theta$-term, a finite baryon chemical potential in QCD, or real-time dynamics are present.
Therefore, alternative first-principles approaches are highly desirable.

The Hamiltonian formulation of lattice gauge theories has attracted increasing attention in recent years, driven by rapid advances in classical tensor network methods and quantum computing.
Since physical states, including the ground state and excited states, can be constructed directly in this formulation, it is well-suited for analyzing systems that suffer from the sign problem in conventional Monte Carlo simulations.
In particular, matrix product states (MPS) provide an efficient tensor network ansatz for constructing optimized physical states in $(1+1)$-dimensional systems.
The $\theta$-term has been incorporated in refs.~\cite{Buyens:2013yza,Buyens:2015tea,Buyens:2017crb,Honda:2022edn,Dempsey:2022nys,Dempsey:2023gib,Itou:2023img,Itou:2024psm,Fujii:2024reh}, while real-time dynamics has been investigated in refs.~\cite{Buyens:2016ecr,Papaefstathiou:2024zsu} for the massive Schwinger model as an application to a nontrivial gauge theory.
Extensions to non-Abelian gauge theories have also been explored~\cite{Banuls:2017ena,Dempsey:2025wia,Dempsey:2026hlf}, 
while finite-density applications within Hamiltonian approaches remain limited, with only our previous MPS study~\cite{Hayata:2023pkw} and the quantum-link-model study of Ref.~\cite{Silvi:2016cas} as notable examples.
A related approach based on the Grassmann tensor renormalization group has been applied to $(1+1)$-dimensional two-color QCD at finite density~\cite{Pai:2024fen}.

In our previous work~\cite{Hayata:2023pkw}, the properties of cold, dense $(1+1)$-dimensional single-flavor $\mathrm{SU}(N_c)$ gauge theory for $N_c=2$ and $N_c=3$ were investigated using MPS on a finite interval with open boundary conditions.
In that study, the gauge field was eliminated by solving the Gauss-law constraint under open boundary conditions, and the resultant Hamiltonian takes the form of a spin system with nonlocal interactions.
The ground state was obtained by minimizing the grand potential at zero temperature using the density-matrix renormalization group method~\cite{2011AnPhy.326...96S,itensor}.
Thermodynamic quantities such as the pressure, the baryon number density, and the equation of state were computed.
That study also discussed how the dynamical degrees of freedom change from hadrons to quarks through the formation of quark
Fermi seas.
One of the remarkable findings is the emergence of an inhomogeneous phase at finite baryon density, where the expectation values of the baryon number density and the chiral condensate are spatially modulated.

However, several important questions remain.
First, it is not yet clear whether the inhomogeneous phase persists in the infinite-volume limit, since spontaneous breaking of a continuous symmetry is disfavored in $(1+1)$-dimensional field theory by the Coleman--Mermin--Wagner theorem~\cite{Mermin:1966fe,Coleman:1973ci}.
Second, a classification of infrared phases at finite baryon density has not yet been established through direct studies based on the Hamiltonian formulation of lattice gauge theory.
Previous studies based on Abelian bosonization under certain gauge-fixing conditions have provided valuable insights, suggesting that, in the strong-coupling limit, a gapped phase is realized for finite quark mass, while a gapless phase appears for vanishing quark mass at zero baryon density~\cite{Baluni:1980bw,Steinhardt:1980ry}.
It has also been argued in ref.~\cite{Lajer:2021kcz} that $(1+1)$-dimensional QCD at finite baryon density can be described, in the long-distance limit, by an effective theory of an interacting fluid known as the Tomonaga--Luttinger liquid~\cite{Tomonaga:1950zz,Luttinger:1960zz,Haldane:1981zza,Haldane:1981zz}.
While these previous works provide important insights into the structure of dense QCD, a direct first-principles verification based on the Hamiltonian formulation of lattice gauge theory has been lacking. 
In particular, the uniform matrix product states (uMPS) framework allows direct access to the thermodynamic limit, which is essential for addressing infrared properties and the fate of inhomogeneous phases.

In the present paper, we directly address these problems by investigating the $(1+1)$-dimensional single-flavor $\mathrm{SU}(2)$ gauge theory at zero and finite baryon density, which we refer to as $\mathrm{QC_2D_2}$, in the thermodynamic limit using uMPS.
In the uMPS framework, physical states are restricted to the translationally invariant states.
This translational invariance significantly reduces the number of variational parameters and eliminates boundary effects.
We construct a gauge invariant ansatz for $\mathrm{QC_2D_2}$ as an extension of the $\mathrm{U}(1)$ gauge theory in uMPS~\cite{Buyens:2013yza}, which is compatible with translational invariance (See also ref.~\cite{Dempsey:2025wia} for the recent study of the gauge invariant uMPS ansatz in the non-Abelian gauge theories).
Ground states are numerically obtained using the variational uniform matrix product state (VUMPS) algorithm~\cite{Zauner-Stauber:2017eqw} with the gauge invariant ansatz.
A key advantage of the present method is that, because the calculation is performed directly in the infinite-volume limit, thermodynamic quantities can be obtained as smooth, continuous functions of the baryon chemical potential.
We also find that the system experiences a phase transition from a gapped phase to a gapless phase when the baryon number density acquires a nonzero expectation value, $ n_B\neq 0$.
At finite density, we perform finite-entanglement scaling to estimate the central charge, and find $c=1$.
Moreover, the two-point correlation function of the baryon number density operator exhibits spatial modulation whose wavenumber is $p=2\pi  n_B \ell$ with $\ell$ being an integer.
From these observations, we conclude that the infrared behavior of $\mathrm{QC_2D_2}$ at finite baryon density is consistent with the Tomonaga--Luttinger liquid theory, which implies that the inhomogeneous phase observed in ref.~\cite{Hayata:2023pkw} does not survive as long-range order in the thermodynamic limit.
The Luttinger parameter is also estimated from the equation of state, and its smooth interpolation from $K \simeq 1$ to $K \simeq 1/N_c$ ($=1/2$ for $N_c=2$) with increasing baryon density constitutes a quantitative signature of the quarkyonic crossover, where $K\simeq 1$ and $K\simeq 1/N_c$ correspond to the baryonic and quark-like regimes, respectively~\cite{McLerran:2007qj,Kojo:2009ha,Kojo:2011cn}.

The paper is organized as follows.
In sec.~\ref{sec:lattice Hamiltonian}, we introduce the lattice-regularized Hamiltonian of $\mathrm{QC_2D_2}$.
In sec.~\ref{sec:gauge-invariant uMPS}, we discuss a manifestly translationally and gauge-invariant uMPS ansatz, as well as properties of the transfer matrix.
In sec.~\ref{sec:results}, we present our numerical results and their physical implications.
Section~\ref{sec:conclusion} is devoted to conclusions and discussion.
In appendix~\ref{appendix:bosonization}, we summarize a brief review and present analytic results for finite-density $\mathrm{QCD}_2$ in the continuum limit.
In appendix~\ref{appendix:local operator on lattice}, we provide the definitions of the local diquark and baryon operators on the lattice.

\section{The Kogut-Susskind formulation of the Hamiltonian gauge theory}\label{sec:lattice Hamiltonian}

In the Hamiltonian formulation of lattice gauge theories, the spatial direction is discretized, while the temporal direction remains continuous.
We consider the $(1+1)$-dimensional $N_c=2$ QCD with a single flavor.
We employ the original Kogut-Susskind Hamiltonian with staggered fermions~\cite{Kogut:1974ag}.
The total Hamiltonian in the temporal gauge is given by 
\begin{align}
    &H_\mathrm{tot}=H_E+H_\mathrm{hop}+H_\mathrm{mass},\label{eq:Hamiltonian SU(2)}\\
    &H_{E}=\sum_n\frac{g^2a}{2}E_a^2(n),\\
    &H_\mathrm{hop}= \sum_n\dfrac{1}{2a}\left(\phi^{\dag c_1}(n+1)U_{c_1}^{c_2}(n)\phi_{c_2}(n)+\mathrm{h.c.}\right),\\
    &H_\mathrm{mass}=m\sum_n(-1)^n\phi^{\dag c}(n)\phi_c(n).
\end{align}
In these expressions, $g$, $m$, $a$, and $n$ denote the bare gauge coupling, the quark mass, the lattice spacing, and the spatial site index, respectively.
$E_a^2(n)$ is the square of the electric field operators, where $a=1,2,3$ is the index of color in the adjoint representation.
$U^{c_2}_{c_1}$ is the link variable in the fundamental representation.
The index $c=1,2$ represents the color degrees of freedom in the fundamental representation for $N_c=2$.
$\phi_c(n)$ and $\phi^{\dag c}(n)$ are the annihilation and creation operators of quarks, respectively.
The continuum limit corresponds to taking $a \to 0$ while keeping the physical coupling fixed.
The Hamiltonian density, which acts on $n$ and $(n+1)$-sites, is given by
\begin{align}
    h(n,n+1)=\dfrac{g^2a}{2}E_a^2(n)+\dfrac{1}{2a}\left(\phi^{\dag c_1}(n+1)U_{c_1}^{c_2}(n)\phi_{c_2}(n)+\mathrm{h.c.}\right)+m(-1)^n\phi^{\dag c}(n)\phi_c(n).
\end{align}
The anti-commutation and commutation relations for the fermions and gauge fields are given by
\begin{align}
    &\{\phi_{c_1}(n),\phi^{\dag c_2}(m)\} =\delta_{nm}\delta_{c_1}^{c_2},\label{eq:fermion anticommutation relation}\\
    &[R_a(n),U(m)]=\delta_{nm}U(m)T_a,\quad [L_a(n),U(m)]=\delta_{nm}T_aU(m).
\end{align}
Here, $R_a(n)$ and $L_a(n)$ are the canonical conjugates of the gauge field on the lattice, which act from the right and left, respectively.
$T_a$ are the generators of $\mathrm{SU}(2)$ in the fundamental representation, which satisfy the following commutation relation, 
\begin{align}
    [T_a,T_b]=i\tensor{f}{_{ab}^{c}}T_c.
\end{align}
Here, $\tensor{f}{_{ab}^{c}}$ denotes the structure constants.
For $N_c = 2$, one has $T_a = \sigma_a/2$, 
where $\sigma_a$ are the Pauli matrices and 
$f_{ab}{}^{c} = \epsilon_{abc}$ is the Levi--Civita symbol. 
Using the Jacobi identity 
\begin{align}
[R_a,[R_b,U]]+[U,[R_a,R_b]]+[R_b,[U,R_a]]=0,
\end{align}
one obtains the following commutation relations,
\begin{align}
    [R_a(n),R_b(m)]=i\delta_{nm}\tensor{f}{_{ab}^{c}}R_c(n),\quad
    [L_a(n),L_b(m)]=-i\delta_{nm}\tensor{f}{_{ab}^{c}}L_c(n).\label{eq:commutation relation of R and L}
\end{align}
Note that the operators $R_a$ and $L_a$ are not independent and are related by parallel transport $R_a=L_b(U^\mathrm{adj})^{b}_a$.
The square of the electric field is given by $E_a^2(n)\coloneqq (L_a)^2(n)=(R_a)^2(n)$.

The Hamiltonian~\eqref{eq:Hamiltonian SU(2)} commutes with the generators of spatial gauge transformations,
\begin{align}
    &G_a(n)=R_a(n)-L_a(n-1)+Q_a(n),\\
    &Q_a(n)=\phi^\dag(n)T_a \phi(n).\label{eq:color charge density}
\end{align}
Here, $Q_a(n)$ denote the color charge density operators.
They obey
\begin{align}
    [Q_a(n),Q_b(m)]=i\delta_{nm}\tensor{f}{_{ab}^{c}}Q_c(n),
\end{align}
and consequently the generators $G_a(n)$ satisfy
\begin{align}
[G_a(n),G_b(m)]=i\delta_{mn}\tensor{f}{_{ab}^c}G_c(n).
\end{align}
The gauge invariance restricts the Hilbert space by the constraint,
\begin{align}
    G_a(n)|\psi\rangle =0.\label{eq:Gauss-law constraint}
\end{align}
This condition is known as the Gauss-law constraint.

We choose the local Hilbert space of the gauge field to be spanned by the states $|j_n, m_{L_n}, m_{R_n}\rangle$, which satisfy
\begin{align}
    &E_a^2|j_n,m_{L_n},m_{R_n}\rangle =j_n(j_n+1)|j_n,m_{L_n},m_{R_n}\rangle,\\
    &R_{3}|j_n,m_{L_n},m_{R_n}\rangle = m_{R_n} |j_n,m_{L_n},m_{R_n}\rangle ,\\
    &L_{3}|j_n,m_{L_n},m_{R_n}\rangle = m_{L_n}|j_n,m_{L_n},m_{R_n}\rangle,\\
    &R_{\pm}|j_n,m_{L_n},m_{R_n}\rangle=\sqrt{j_n(j_n+1)-m_{R_n}(m_{R_n}\pm 1)}|j_n,m_{L_n},m_{R_n}\pm 1\rangle,\\
    &L_{\pm}|j_n,m_{L_n},m_{R_n}\rangle=\sqrt{j_n(j_n+1)-m_{L_n}(m_{L_n}\mp 1)}|j_n,m_{L_n}\mp 1,m_{R_n}\rangle .\label{eq:action_Lpm}
\end{align}
Here, we define $R_{\pm}\coloneqq R_1\pm i R_2$ and $L_{\pm}\coloneqq L_1\pm i L_2$. 
The states satisfy the orthonormality condition, 
\begin{align}
\langle j_1,m_{L_1},m_{R_1}|j_2,m_{L_2},m_{R_2}\rangle =\delta_{j_1j_2}\delta_{m_{L_1}m_{L_2}}\delta_{m_{R_1}m_{R_2}}.
\end{align}
The spin quantum number $j_n$ takes integer or half-integer values.
The magnetic quantum numbers $m_{L_n}$ and $m_{R_n}$ run from $-j_n$ to $j_n$ in integer steps.
Note that the opposite ladder actions of $R_{\pm}$ and $L_{\pm}$ 
arise from the sign difference in their commutation relations~\eqref{eq:commutation relation of R and L}.

Let us next discuss the local Hilbert space for the fermion.
Due to Fermi statistics, there are only four possible states:
\begin{align}
    |0_n\rangle,\quad|c_n\rangle\equiv \phi^{\dag c}|0_n\rangle,\quad|B_n\rangle\equiv \frac{\epsilon_{c_1c_2}}{2}\phi^{\dag c_1}\phi^{\dag c_2}|0_n\rangle.
\end{align}
Here, $|0_n\rangle$ is the Fock vacuum defined by the relation $\phi_c(n)|0_n\rangle=0$.
Applying the $Q_a(n)$ operator to the above states, we can find that $|0_n\rangle$ and $|B_n\rangle$ states are singlet states, while $|c_n\rangle$ is the doublet state.
It is convenient to label these states by $|q_n,j_{M_n},m_{M_n}\rangle$, where $q_n,~j_{M_n}$ and $m_{M_n}$ are the quark occupancy number, the spin quantum number for $Q_a(n)$, and the eigenvalue of $Q_{a=3}(n)$, respectively.
In particular, one obtains the one-to-one correspondence,
\begin{align}
    &|0_n\rangle =|q_n=0,j_{M_n}=0,m_{M_n}=0\rangle,\nonumber\\
    &|c_n=1\rangle=|q_n=1,j_{M_n}=1/2,m_{M_n}=1/2\rangle,\quad |c_n=2\rangle=|q_n=1,j_{M_n}=1/2,m_{M_n}=-1/2\rangle,\label{eq:fermion basis}\\
    &|B_n\rangle =|q_n=2,j_{M_n}=0,m_{M_n}=0\rangle.\nonumber
\end{align}
The basis $|q_n,j_{M_n},m_{M_n}\rangle$ is more convenient for constructing gauge-invariant states, as we shall see in the next section.

\section{Gauge invariant uniform matrix product state}\label{sec:gauge-invariant uMPS}
In this section, we introduce a gauge-invariant uMPS ansatz and discuss the properties of the corresponding transfer matrices.
Before introducing the uMPS ansatz, let us clarify how the Hilbert space is decomposed into local degrees of freedom.
In lattice gauge theory, the gauge and matter fields reside on links and sites, and the physical Hilbert space is defined by the Gauss-law constraint~\eqref{eq:Gauss-law constraint}.
Therefore, the physical Hilbert space does not factorize into a simple tensor product.

To construct an MPS, it is convenient to reorganize the Hilbert space into site-local building blocks.
This is possible because the Hilbert space of gauge fields on a link can be decomposed as
\begin{equation}
\mathcal H_{\text{link}(n)}
=
\bigoplus_{j_n}
\left(
\mathcal{H}_{j_n}^{(L)}
\otimes
\mathcal{H}_{j_n}^{(R)}
\right) \subset \mathcal{H}_{L(n)}\otimes \mathcal{H}_{R(n)},
\label{eq:decomposition_Hilbertspace}
\end{equation}
where $\mathcal{H}_{L(n)}=\oplus_{j_n} \mathcal{H}^{(L)}_{j_n}$ and $\mathcal{H}_{R(n)}=\oplus_{j_n}\mathcal{H}^{(R)}_{j_n}$. 
Here, $\mathcal{H}_{j_n}^{(L)}$ and $\mathcal{H}_{j_n}^{(R)}$ carry
irreducible representations of $\mathrm{SU}(2)$ generated by $L_a(n)$ and
$R_a(n)$, respectively.
The basis states $\mathcal H_{\text{link}(n)}\owns|j_n,m_{L_n},m_{R_n}\rangle=|j_n,m_{L_n}\rangle_L\otimes |j_n,m_{R_n}\rangle_R\in \mathcal{H}_{j_n}^{(L)}\otimes\mathcal{H}_{j_n}^{(R)}$ explicitly realize this left--right factorization.

Using this structure, we associate the left electric degrees of freedom
of link $(n-1)$ and the right electric degrees of freedom of link $(n)$
with site $n$.
We thus introduce the extended local Hilbert space at site $n$ as
\begin{align}
\mathcal{H}_{(n)}^{\mathrm{ext}}
\coloneqq
\mathcal{H}_{L(n-1)}\otimes\mathcal{H}_{M(n)}\otimes\mathcal{H}_{R(n)},
\label{eq:extended_Hilbertspace}
\end{align}
where $\mathcal{H}_{M(n)}$ denotes the fermionic Hilbert space at site $n$.
A basis state $|s_n\rangle\in \mathcal H_{(n)}^{\mathrm{ext}}$
is labeled by
\begin{align}
s_n =
\big(
j_{L_{n-1}},m_{L_{n-1}};
q_n,j_{M_n},m_{M_n};
j_{R_n},m_{R_n}
\big).
\end{align}
Within this extended local space, the Gauss-law constraint becomes a
purely local condition.
The Gauss-law constraint can be solved by using the Clebsch-Gordan coefficients $C^{J,M}_{j_1,m_1,j_2,m_2}$  as
\begin{align}
 \sum_{m_{R_n},m_{L_{n-1}},m_{M_n}}C^{j_{L_{n-1}},m_{L_{n-1}}}_{j_{R_{n}},m_{R_{n}},j_{M_{n}},m_{M_{n}}}
|j_{L_{n-1}},m_{L_{n-1}}\rangle_L\otimes|q_n,j_{M_n},m_{M_n}\rangle\otimes|j_{R_n},m_{R_n}\rangle_R.
\label{eq:gauge_inv_local_state}
\end{align}
One can directly verify the Gauss-law constraint by acting $G_a(n)$ on the state in eq.~\eqref{eq:gauge_inv_local_state}.
Although the Gauss-law constraint is now implemented locally,
the total Hilbert space $\bigotimes_n \mathcal H_{(n)}^{\mathrm{ext}}$ is larger than the original physical Hilbert space (see eqs.~\eqref{eq:decomposition_Hilbertspace} and \eqref{eq:extended_Hilbertspace}).
This is because the left and right electric representations on each link are treated as independent degrees of freedom.
To recover the original physical Hilbert space, one must additionally impose
the consistency condition that the $\mathrm{SU}(2)$ representations on
the right of one site and on the left of its neighbor coincide.
As we will see below, within the MPS framework, this constraint is naturally enforced by restricting the bond space to fixed representation sectors and by contracting only matching blocks.

Let us construct the gauge-invariant uMPS. We assume that the ground state is invariant under translations by two
lattice sites, since a translation by one lattice spacing corresponds
to a discrete chiral transformation in the staggered-fermion
formulation.
We therefore employ a two-site translationally invariant uMPS.
The ansatz is given by
\begin{equation}
|\Psi_\mathrm{uMPS}\rangle
=
\sum_{{\bm s}}\cdots A^{s_{2r}}B^{s_{2r+1}}A^{s_{2r+2}}B^{s_{2r+3}} \cdots
|\cdots s_{2r},s_{2r+1},s_{2r+2},s_{2r+3}\cdots\rangle ,
\label{eq:two_site_umps}
\end{equation}
where $A$ and $B$ are MPS tensors associated with the even and odd
lattice sites, respectively.
The summation is taken over the entire local Hilbert space, ${\bm s}=(\cdots,s_{2r},s_{2r+1},s_{2r+2},s_{2r+3},\cdots)$. 

The gauge-invariant uMPS ansatz was introduced in ref.~\cite{Buyens:2013yza} for $\mathrm{U}(1)$ gauge theory.
As a natural extension to $\mathrm{SU}(2)$ gauge theories, we find the following MPS ansatz that is manifestly gauge invariant,
\begin{align}
    [A^{s_{2r}}]_{J_L\alpha_{J_L};J_R\beta_{J_R}}
    &={a}^{j_{L_{2r-1}},j_{M_{2r}},q_{2r},j_{R_{2r}}}_{\alpha_{J_L}\beta_{J_R}}
    \frac{1}{\sqrt{d_{j_{L_{2r-1}}}}}C^{j_{L_{2r-1}},m_{L_{2r-1}}}_{j_{R_{2r}},m_{R_{2r}},j_{M_{2r}},m_{M_{2r}}}\delta^{j_{L_{2r-1}}}_{J_L}\delta^{j_{R_{2r}}}_{J_{R}},\\
    [B^{s_{2r+1}}]_{J_L\alpha_{J_L};J_R\beta_{J_R}}
    &={b}^{j_{L_{2r}},j_{M_{2r+1}},q_{2r+1},j_{R_{2r+1}}}_{\alpha_{J_L}\beta_{J_R}}
    \frac{1}{\sqrt{d_{j_{L_{2r}}}}}C^{j_{L_{2r}},m_{L_{2r}}}_{j_{R_{2r+1}},m_{R_{2r+1}},j_{M_{2r+1}},m_{M_{2r+1}}}\delta^{j_{L_{2r}}}_{J_L}\delta^{j_{R_{2r+1}}}_{J_{R}}.
\end{align}
Here, $d_{j}=2j+1$ is the dimension of the spin-$j$ representation.
The variational parameters ${a}^{j_{L_{2r-1}},j_{M_{2r}},q_{2r},j_{R_{2r}}}_{\alpha_{J_L}\beta_{J_R}}$ and ${b}^{j_{L_{2r}},j_{M_{2r+1}},q_{2r+1},j_{R_{2r+1}}}_{\alpha_{J_L}\beta_{J_R}}$ are the gauge-invariant variational degrees of freedom.\footnote{Since the model possesses a global $\mathrm{U}(1)$ symmetry corresponding to baryon number conservation, one may exploit this symmetry to further reduce the number of variational degrees of freedom. In the present work, we do not exploit this symmetry for simplicity and leave it as a future study.}
$\alpha_j=1,2,\cdots,D_{j}$ is the $j$-dependent bond dimension,
whose total dimension is given by $D=\sum^{j_\mathrm{cutoff}}_{j=0}D_j$, where $j_\mathrm{cutoff}$ is the cutoff of the spin imposed in the simulation.
The factors $\delta^{j_{L_{2r-1}}}_{J_L}\delta^{j_{R_{2r}}}_{J_R}$ and $\delta^{j_{L_{2r}}}_{J_L}\delta^{j_{R_{2r+1}}}_{J_{R}}$
identify the virtual bond representations with the physical electric representations.
As neighboring tensors are contracted, this identification automatically
enforces representation matching on each link ($j_{R_n}=j_{L_n}$).

In this work, we implement a single-site uMPS, where even and odd sites are combined and treated as a single site.
Our uMPS ansatz is therefore invariant under translations by one site and is explicitly given by
\begin{align}
    &[\mathcal{C}^{s_{2r},s_{2r+1}}]_{J_L \alpha_{J_L};J_R\beta_{J_R}}:=[A^{s_{2r}}{B}^{s_{2r+1}}]_{J_L\alpha_{J_L};J_R\beta_{J_R}}\nonumber\\
    &=c_{\alpha_{J_L}\beta_{J_R}}^{j_{L_{2r-1}},j_{M_{2r}},q_{2r},j_I,j_{M_{2r+1}},q_{2r+1},j_{R_{2r+1}}}
\frac{1}{\sqrt{d_{j_{L_{2r-1}}}}}C^{j_{L_{2r-1}},m_{L_{2r-1}}}_{j_{I},m_{R_{2r}},j_{M_{2r}},m_{M_{2r}}}
\frac{1}{\sqrt{d_{j_{I}}}}C^{j_{I},m_{L_{2r}}}_{j_{R_{2r+1}},m_{R_{2r+1}},j_{M_{2r+1}},m_{M_{2r+1}}}
\delta^{j_{L_{2r-1}}}_{J_L}\delta^{j_{R_{2r+1}}}_{J_{R}},\label{eq:gauge-invariant ansatz}
\end{align}
where 
\begin{align}
c_{\alpha_{J_L}\beta_{J_R}}^{j_{L_{2r-1}},j_{M_{2r}},q_{2r},j_I,j_{M_{2r+1}},q_{2r+1},j_{R_{2r+1}}}
=
\sum_{\alpha_{J_I}}{a}^{j_{L_{2r-1}},j_{M_{2r}},q_{2r},j_{I}}_{\alpha_{J_L}\alpha_{J_I}}{b}^{j_{I},j_{M_{2r+1}},q_{2r+1},j_{R_{2r+1}}}_{\alpha_{J_I}\beta_{J_R}}.
\end{align}
One should note that the gauge field cannot be eliminated in the uMPS, unlike in finite systems with open boundary conditions.

Because of the block diagonal structure of the bond space, Schmidt coefficients are labeled by the spin $j$ as 
\begin{align}
    |\psi\rangle = \sum_{j=0,1/2,\cdots}^{j=j_\mathrm{cutoff}}\sum^{D_j}_{\alpha_j=1}\sqrt{\sigma_{\alpha_j}}|\psi_{j\alpha_j L}\rangle\otimes|\psi_{j\alpha_j R}\rangle,~\sum_{j=0,1/2,\cdots}^{j_\mathrm{cutoff}}\sum^{D_{j}}_{\alpha_j=1}\sigma_{\alpha_j}=1.
\end{align}
In this expression, $\sigma_{\alpha_j}$ denote the squared Schmidt coefficients of the $j$ sector, which is assumed to be in descending order ($\sigma_{\alpha_j=1}\geq\sigma_{\alpha_j=2}\geq\cdots\geq\sigma_{\alpha_j=D_j}$).
Because the contribution of the electric field to the energy density is proportional to $j(j+1)$, Schmidt coefficients of the higher sectors are increasingly suppressed.
Indeed, we observe that larger values of $j$ lead to a strong suppression of $\sigma_{j}$ for a finite lattice spacing.
Since the distribution of the Schmidt coefficients is highly nontrivial, we control the truncation error $\varepsilon$ instead of fixing the bond dimension directly in the simulation.
More concretely, the bond dimensions $D_j$ and the cutoff $j_\mathrm{cutoff}$ are chosen such that 
\begin{align}
    \begin{cases}
    \sigma_{\alpha_{j}=D_j+1}<\varepsilon<\sigma_{\alpha_j=D_j},~(0\leq j\leq j_\mathrm{cutoff})\\
    \sigma_{\alpha_{j}=1}<\varepsilon,~(j_\mathrm{cutoff}<j)
    \end{cases}
\end{align}
are satisfied.
In the simulations, we monitor the Schmidt coefficients at each iteration of VUMPS to dynamically control $j_\mathrm{cutoff}$ and $D_j$ so that the above conditions are maintained.
In summary, the input parameters of our simulations are $ga,~m/g,~\mu_B/g$, and $\varepsilon$.
The outputs of the simulations are $j_\mathrm{cutoff}$, $D_j$, and variational parameters $c_{\alpha_{J_L}\beta_{J_R}}^{j_{L_{2r-1}},j_{M_{2r}},q_{2r},j_I,j_{M_{2r+1}},q_{2r+1},j_{R_{2r+1}}}$.
The total bond dimension is given by $D=\sum^{j_\mathrm{cutoff}}_{j=0}D_j$.

\begin{figure*}[htbp]
    \centering
    \begin{minipage}[t]{0.45\textwidth}
        \centering
        \includegraphics[width=\textwidth]{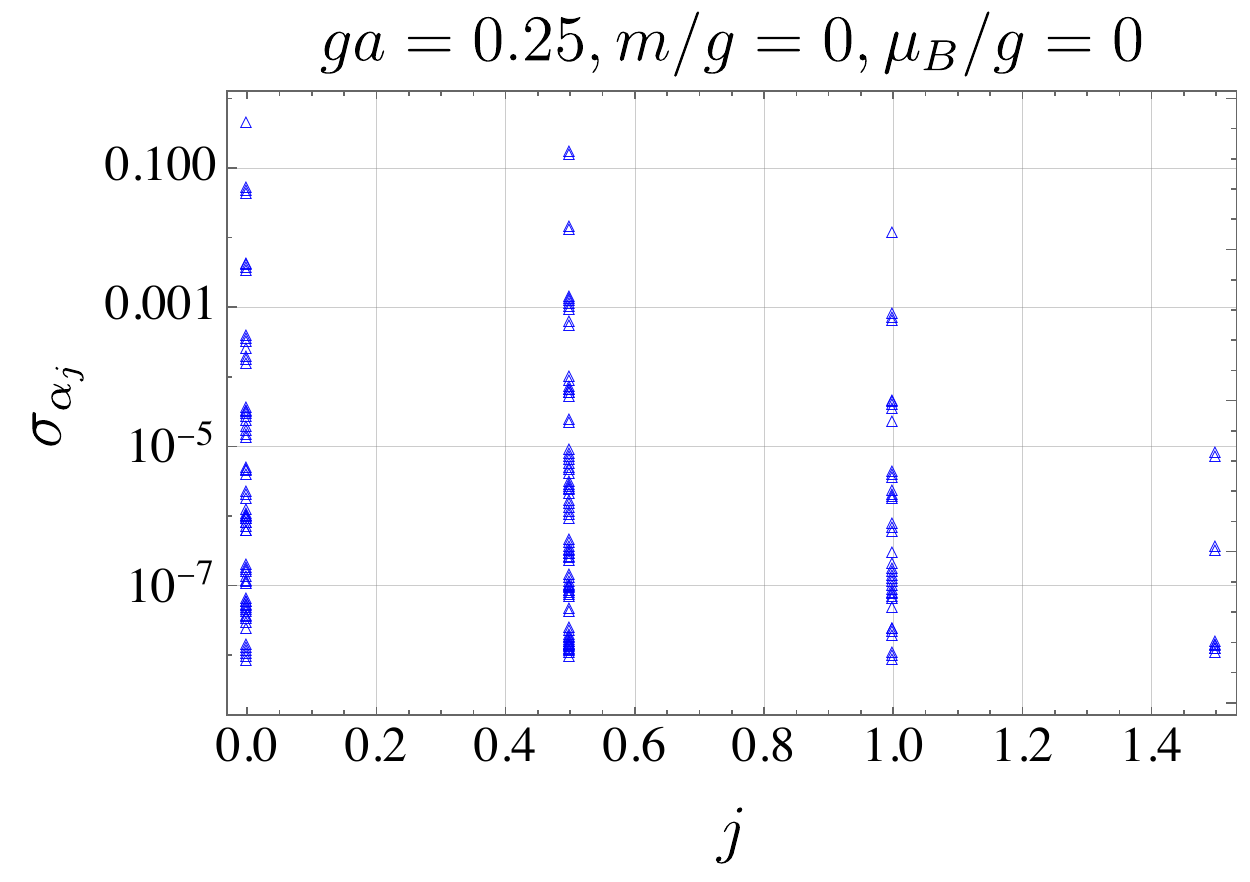}
    \end{minipage}
    \hspace{0.01\textwidth} 
    \begin{minipage}[t]{0.45\textwidth}
        \centering
        \includegraphics[width=\textwidth]{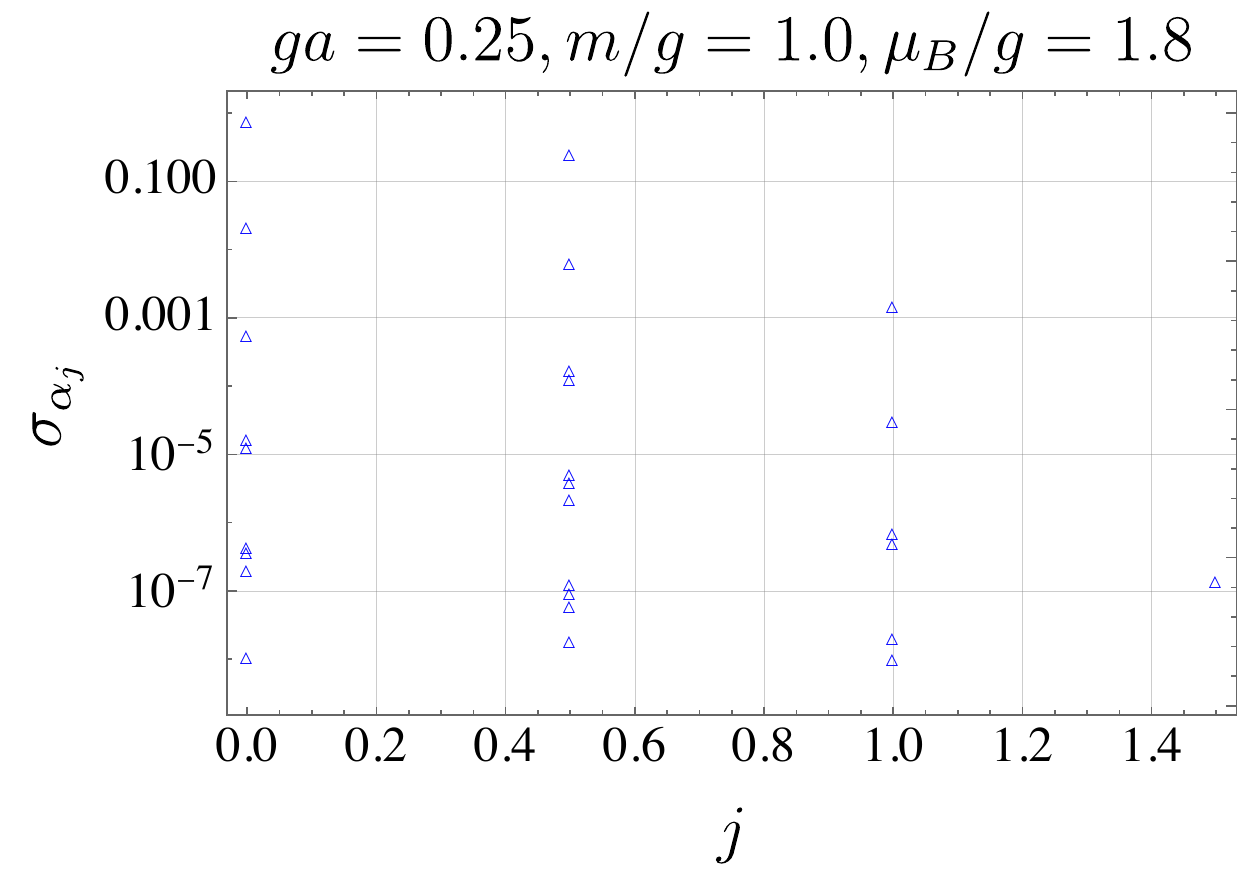}
    \end{minipage}
\caption{
Dependence of the Schmidt spectrum on the representation $j$ is shown for the critical case ($ga = 0.25$, $m/g = 0$, $\mu_B = 0$; left) and for the gapped case ($ga = 0.25$, $m/g = 1.0$, $\mu_B/g = 1.8$; right).
}\label{fig:Schmidt}
\end{figure*}
Figure \ref{fig:Schmidt} shows the Schmidt spectra $\sigma_{\alpha_j}$ as a function of the representation $j$.
As we shall discuss later, the massless point $m/g=0$ with a finite lattice spacing $ga\neq 0$ behaves as a critical point, while a finite quark mass with vanishing baryon number density $ n_B=0$ is considered to be a gapped phase.
It is clear from the figure that Schmidt coefficients in the higher spin sector are exponentially suppressed in both cases.
Therefore, it is sufficient to truncate the spin sector at $j_\mathrm{cutoff}=3/2$ for $ga =0.25$ and $\varepsilon\geq 10^{-8}$.
For a smaller lattice spacing or truncation error, the $j=2$ sector is no longer negligible.
This observation is consistent with the previous study on a finite interval with an open boundary condition~\cite{Banuls:2017ena}.

We next discuss some basic properties of the transfer matrices of ground states.
The transfer matrix is defined by
\begin{align}
    (T_{\mathcal{C}})_{\alpha_1\alpha_3;\alpha_2\alpha_4}=\sum_{\kappa}\mathcal{C}^{\kappa}_{{\alpha_1 \alpha_2}}(\mathcal{C}^{\kappa})^*_{{\alpha_3\alpha_4}}.
\end{align}
Here, $\mathcal{C}^{\kappa}_{\alpha_1\alpha_2}$ is the local MPS tensor defined by~\eqref{eq:gauge-invariant ansatz}, where $\kappa=(s_{2r},s_{2r+1})$ and $\alpha=(J_L\alpha_{J_L};J_R\beta_{J_R})$ collectively represent the local Hilbert space and bond index, respectively.
We assume that the transfer matrix admits the following spectral decomposition,
\begin{align}
    T_{\mathcal{C}}=\sum_{k=0}^{D^2-1}\lambda_k|k)(k|,\quad |\lambda_0|=1>|\lambda_1|\geq\cdots\geq|\lambda_{D^2-1}|.\label{eq:eigendecomposition}
\end{align}
Here, $\lambda_k$ are the eigenvalues of $T_{\mathcal{C}}$; $|k)$ and $(k|$ denote the corresponding right and left eigenvectors, respectively.
The largest eigenvalue is given by $\lambda_{k=0}=1$ by the construction of uMPS, and is assumed to be non-degenerate.
The left and right eigenvectors are normalized such that $(k|k')=\delta_{k,k'}$.
The two-point function of two local operators $O_{i_1}(n_1)$ and $O_{i_2}(n_2)$ can be expressed as
\begin{align}
    &\langle\Psi_\mathrm{uMPS}|O_{i_1}(2r_1,2r_1+1)O_{i_2}(2r_2,2r_2+1)|\Psi_\mathrm{uMPS}\rangle=\sum_{k=0}Z_{k,i_1}Z_{k,i_2}e^{-\epsilon_k|r_2-r_1-1|+i\phi_k|r_2-r_1-1|},\label{eq:two-point function in uMPS}\\
    &Z_{k,i_1}=\sum_{\kappa_1,\kappa_2}[O_{i_1}(2r_1,2r_1+1)]_{\kappa_2\kappa_1}
    (0|\mathcal{C}^{\kappa_1}(\mathcal{C}^{\kappa_2})^*|k),~Z_{k,i_2}=\sum_{\kappa_1,\kappa_2}[O_{i_2}(2r_2,2r_2+1)]_{\kappa_2\kappa_1}
    (k|\mathcal{C}^{\kappa_1}(\mathcal{C}^{\kappa_2})^*|0),\label{eq:correlation function uMPS}
\end{align}
where $\epsilon_k=-\log|\lambda_k|$ and $\phi_k=\arg\lambda_k$. 
The coefficients $Z_{k,i_1}$ and $Z_{k,i_2}$ are form factors determined by the matrix elements of the local operators,
with $[O_{i}(2r,2r+1)]_{\kappa_2\kappa_1}=\langle\kappa_2|O_i(2r,2r+1)|\kappa_1\rangle$.

In this calculation, we use eq.~\eqref{eq:eigendecomposition}.
Therefore, the correlation length at the long-distance limit can be extracted from its asymptotic behavior,
\begin{align}
    \langle\Psi_\mathrm{uMPS}|O_{i_1}(2r_1,2r_1+1)O_{i_2}(2r_2,2r_2+1)|\Psi_\mathrm{uMPS}\rangle\sim Z_{k_m,i_1}Z_{k_m,i_2}e^{-\epsilon_{k_m}|r_2-r_1-1|+i\phi_{k_m}|r_2-r_1-1|},\quad(|r_2-r_1|\to \infty). \label{eq:lattice version of two-point function}
\end{align}
Here, $k_m > 0$ denotes the smallest index for which the form factors are nonzero, i.e., $Z_{k_m,i_1} \neq 0$ and $Z_{k_m,i_2} \neq 0$.
This implies that the correlation length can be extracted directly from the transfer matrix spectrum, without fitting the two-point function, thanks to translational invariance. 
In particular, the longest correlation length for the spatial direction is given by $\xi =\epsilon_{k_m}^{-1}$ in lattice units.
In the continuum limit, where Lorentz invariance is restored, the transfer matrix $T_{\mathcal{C}}$ can be interpreted as an imaginary-time evolution operator.
Consequently, $\xi$ can be identified as the inverse of the mass gap as long as the corresponding form factor is nonzero~\cite{Zauner:2014iea}.
Although $\epsilon_{k=1}$ approaches zero in a critical phase as we increase the bond dimension, it is obvious that a nonvanishing correlation length is inevitably induced, $\epsilon_{k=1}>0$ in uMPS realization.
When the two-point function exhibits spatial modulations, their wavenumbers can be extracted from $\phi_{k_m}$.
We will use $\phi_{k_m}$ to identify the spatial modulations of two-point functions that appear at finite density.

A linear combination of eigenvalues of the transfer matrix defined by
\begin{align}
    \delta=\sum_i c_i\epsilon_i,~\sum_ic_i=0,\label{eq:delta}
\end{align}
quantifies the discreteness of continuous exponential spectra of the correlation function in eq.~\eqref{eq:correlation function uMPS} induced by the finite bond dimension.
In the limit of $\varepsilon\to 0$ ($D\to \infty$), $\delta$ should approach zero such that the sum of exponential spectra reproduces the power law dependence of correlation functions~\cite{Zauner:2014iea}.
Interestingly, it is found in ref.~\cite{Rams:2018uwo} that $\delta$ is a better refinement parameter than the bond dimension to perform the extrapolation of $\varepsilon\to 0$ ($D\to \infty$) corresponding to $\delta \to 0$.
This observation is particularly important to extract precise correlation lengths because the convergence of $\epsilon_j$ ($j\ne  0$) is not fast enough, even in the gapped phase.
We perform this extrapolation method to extract a baryon mass.

\section{Results}\label{sec:results}

In this section, we present numerical results obtained from gauge-invariant optimized ground states using the VUMPS algorithm at zero and finite baryon densities.
To determine the zero-temperature equilibrium state at a fixed baryon chemical potential $\mu_B$, we minimize
\begin{align}
\langle\Psi_\mathrm{uMPS}|H_\mathrm{tot}-\mu_BN_B|\Psi_\mathrm{uMPS}\rangle, \label{eq:target function}
\end{align}
where $N_B$ is the baryon number operator.
Here, the baryon chemical potential $\mu_B$ and the  baryon number operator $N_B$ are related through
\begin{align}
N_q=\sum_n j^0_q(n)=\sum_n\phi^\dag(n)\phi(n)=N_cN_B
\end{align}
and $\mu_B=N_c\mu_q$.
The energy and quark number density operators are given by 
\begin{align}
h=(h(n,n+1)+h(n+1,n+2))/2,\quad j^0_q=(j^0_q(n)+j^0_q(n+1))/2.
\end{align}
Our numerical simulations find that there is a phase transition between zero and nonzero baryon densities.
Since these two phases are qualitatively different, we discuss their properties separately in the following subsections.

\subsection{Vacuum properties}

In this subsection, we show our numerical results at zero baryon chemical potential, $\mu_B=0$.
As mentioned in the introduction, some exact results can be obtained in the strong coupling regime, $g\gg m$,
using the bosonization technique.
We provide a brief review in appendix~\ref{appendix:bosonization}.
For $0<m/g\ll1$, the system is described by the sine-Gordon model, whose excitations consist of kinks (and anti-kinks) and their bound states, known as breathers~\cite{Steinhardt:1980ry}.
At the bare massless point $m=0$, the system turns into a critical phase and is described by a single free compact boson corresponding to the baryon number density mode.
Consequently, a conformal field theory with central charge $c=1$ emerges.

We first study the gapped phase realized for nonvanishing quark masses $m/g=0.5$ and $m/g=1$ at fixed lattice spacing $ga=0.25$.
We evaluate the eigenvalues of the transfer matrix from constructed ground states.
The longest correlation length in lattice units is identified as $\epsilon_1^{-1}$, where $\epsilon_1 = -\log|\lambda_1|$ with $\lambda_1$ being the second largest eigenvalue of the transfer matrix (see eq.~\eqref{eq:two-point function in uMPS}).
Interestingly, we find a degenerate spectrum, $\epsilon_1\simeq \epsilon_2\simeq\epsilon_3$ which suggests three degenerate excited states.
As discussed in Appendix~\ref{appendix:bosonization}, a single-flavor $\mathrm{QC_2D_2}$ has an $\mathrm{SU}(2)/\mathbb{Z}_2\simeq \mathrm{SO}(3)$ flavor symmetry.
The triplet spectrum may reflect the presence of this symmetry.

To determine the baryonic quantum numbers of the triplet excitations, we compute the form factors of the scalar diquark and anti-diquark operators defined in Appendix~\ref{appendix:local operator on lattice}.
We find that two of the three degenerate excitations have nonzero form factors, while the remaining one has a form factor consistent with zero within numerical precision.
This suggests that the former two excitations carry baryon and anti-baryon quantum numbers, whereas the latter is baryon-number neutral and does not couple to these operators.
From this observation, we interpret the triplet as being composed of a baryon, an anti-baryon, and a non-baryonic state.\footnote{
More precisely, for each degenerate triplet, two eigenvalues corresponding to baryonic and anti-baryonic states agree within machine precision, whereas the third eigenvalue corresponding to a non-baryonic state is slightly split, with a relative deviation of order $10^{-3}$ for $ga=0.25$ and $m/g\gtrsim 0.5$.
The degeneracy of two eigenvalues within machine precision may be related to an exact charge conjugation symmetry on the lattice.
Because the $\mathrm{SU}(2)/\mathbb{Z}_2$ flavor symmetry is explicitly broken on the lattice with staggered fermions, the observed tiny deviation of degeneracy of the third spectrum may be caused by the finite lattice spacing effect.
We also simulate at a large lattice spacing $ga=2.0$ and $m/g=1.0$, and find that there are only doublet and singlet spectra but no triplet spectra in the low-lying mode, which indirectly supports our claim.
}
In the strong coupling limit $g\gg m$, the triplet is composed of the kink (baryon), anti-kink (anti-baryon), and the breather (meson) states~\cite{Steinhardt:1980ry}.

We identify the baryon mass with the inverse of physical correlation length, $M_B/g=\epsilon_1(2ga)^{-1}$.
As we discussed around eq.~\eqref{eq:delta}, a precise estimate of $\epsilon_1$ requires an extrapolation procedure.
Regarding the degeneracy in the spectrum, we introduce $\delta=\epsilon_4-\epsilon_1$ and perform a linear fit of the form
$\epsilon_1=a_{\delta} \delta + \epsilon_{\infty}$, where $a_\delta$ and $\epsilon_{\infty}$ are fitting parameters.
From this extrapolation, we obtain improved values of baryon mass $M_B/g\simeq 2.32$ for $m/g=1.0$ and $M_B/g\simeq 1.34$ for $m/g=0.5$.
These values coincide with the threshold values of the baryon chemical potential that induces the nonzero baryon number, as we will see in the next subsection.

We next study the infrared behavior of this theory at the quark massless point $m=0$.
We find that both the correlation length and entanglement entropy diverge as the truncation error is reduced.
This simultaneous growth is consistent with a gapless, critical phase.
The entanglement entropy between two semi-infinite subsystems separated at a given point is computed from its definition, 
\begin{align}
S_\mathrm{EE}=-\sum_{j}\sum^{D_j}_{\alpha_j=1}\sigma_{\alpha_j}\log\sigma_{\alpha_j}.
\end{align}
The central charge can be extracted from the well-known relation
$S_\mathrm{EE}\simeq (c/6)\log(\xi/a)$ in conformal field theory \cite{Calabrese:2004eu,Calabrese:2009qy},
where $\xi,~a$ and $c$ are the correlation length in the physical unit, the lattice spacing, and the central charge, respectively.
This expression is valid in the regime $\xi\gg a$, and the nonuniversal constant term independent of $\xi$ has been omitted.
However, it is difficult to determine the central charge directly from this formula, because the convergence of the correlation length becomes worse as the system approaches the critical point.
For this reason, the finite-entanglement scaling method~\cite{Pollmann:2009lnv} is widely used to extract the central charge in the MPS calculations.
The main idea is that the finite correlation length induced by a finite bond dimension, $\epsilon_1^{-1}(D)$ is identified as a perturbation away from criticality (see ref.~\cite{Schneider:2024ipg} for a further in-depth discussion).
Assuming the empirical scaling relation $\epsilon^{-1}_1(D)\propto D^\kappa$ near the critical point, one can extract the central charge by fitting the entanglement entropy to $S_\mathrm{EE}\simeq(c\kappa/6)\log(D)$.

\begin{figure*}[htbp]
    \centering
    \begin{minipage}[t]{0.45\textwidth}
        \centering
        \includegraphics[width=\textwidth]{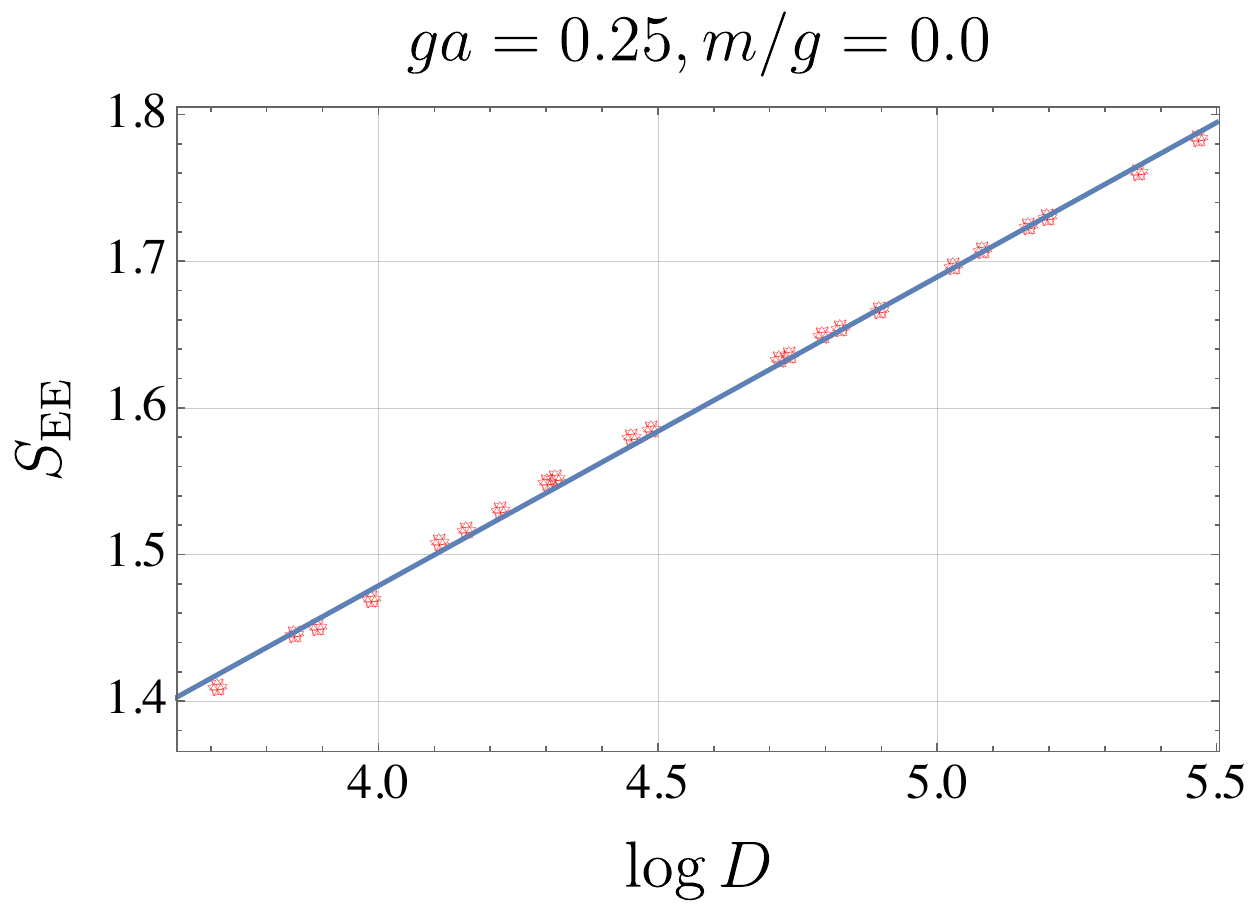}
    \end{minipage}
    \hspace{0.01\textwidth} 
    \begin{minipage}[t]{0.45\textwidth}
        \centering
        \includegraphics[width=\textwidth]{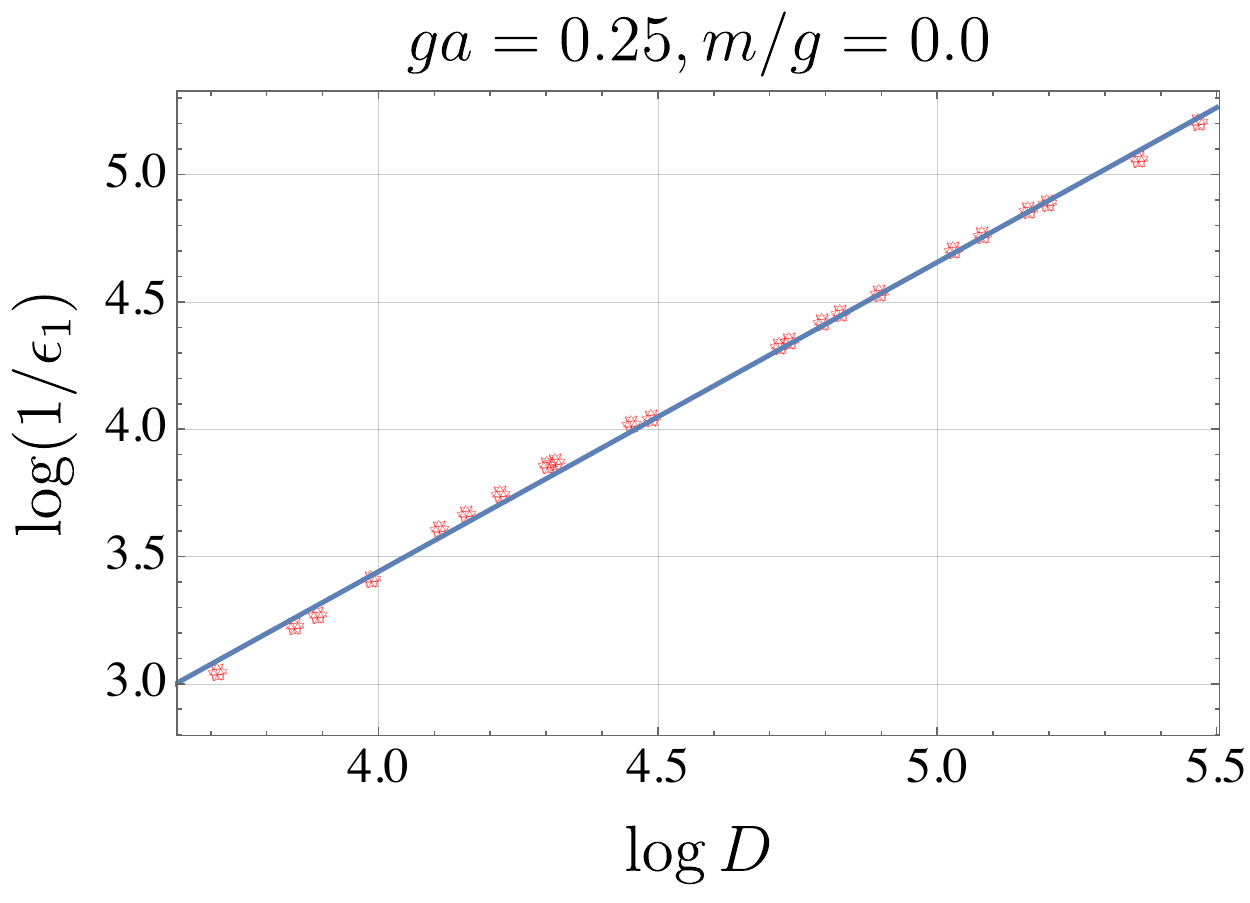}
    \end{minipage}
\caption{
Dependence of the entanglement entropy $S_{\rm EE}$ (left) and the correlation length
$\xi=1/\epsilon_1$ in lattice units (right) on the total bond dimension $D$.
Data points and linear fits are shown.
}\label{fig:entanglement scaling}
\end{figure*}

Figure~\ref{fig:entanglement scaling} shows entanglement entropy and the correlation length in lattice units as functions of the total bond dimension $D$ for fixed lattice parameters $ga=0.25$ and $m/g=0$.
One can clearly see the empirical scaling property $\log(1/\epsilon_1)\propto\kappa\log D$. 
The entanglement entropy obeys the scaling relation $S_\mathrm{EE}\propto \log D$.
Once $\kappa$ is determined from a fit of $\log \xi$ versus $\log D$, the central charge can be extracted according to ref.~\cite{Stojevic:2014zta}.
Using this procedure, the best-fit value is found to be
\begin{align}
c_\mathrm{IR}=1.04.
\end{align}
To the best of our knowledge, this is the first determination of the central charge for $\mathrm{SU}(2)$ gauge theory with a single fundamental fermion at $m/g = 0$ based on a first-principles lattice calculation.\footnote{A similar analysis can be performed in the limit $ga \to 0$ with fixed $m/g \neq 0$. 
In this limit, the gauge coupling $g$ effectively vanishes and the theory reduces to $N_c = 2$ free massless Dirac fermions, each contributing $c = 1$, giving $c_\mathrm{UV} = 2$. This is consistent with the previous study~\cite{Banuls:2017ena}.}

\subsection{Properties at finite baryon density}

In this subsection, we present our numerical results at finite density.
We show that the gapless phase realized for $\mu_B > M_B$ is described by a Tomonaga--Luttinger liquid with central charge $c=1$. Its Luttinger parameter decreases from $K\simeq 1$ to $K\simeq 1/2$ with increasing density, and the emergence of a quark Fermi sea is consistent with the quarkyonic picture.

\subsubsection{Thermodynamic properties}
The energy density, the baryon number density, and the pressure in physical units are defined as 
\begin{align}
    \ed =\langle\Psi_\mathrm{uMPS}|h|\Psi_\mathrm{uMPS}\rangle/a,\quad
    n_B=\langle\Psi_\mathrm{uMPS}|j^0_B|\Psi_\mathrm{uMPS}\rangle/a,\quad
    P=\mu_Bn_B-\ed,
\end{align}
where $j^0_B=j^0_q/N_c$.
These quantities are computed from the optimized ground states obtained using the VUMPS algorithm.
In our simulations, we first construct the ground state with a relatively large truncation error, $\varepsilon = 10^{-4}$, 
and then gradually improve the accuracy to $\varepsilon = 10^{-8}$ for fixed lattice parameters $ga$, $m/g$, and $\mu_B/g$.
We confirm that the results, except for the onset of nonzero baryon number density, are insensitive to the truncation error.

\begin{figure*}[htbp]
    \centering
    \begin{minipage}[t]{0.45\textwidth}
        \centering
        \includegraphics[width=\textwidth]{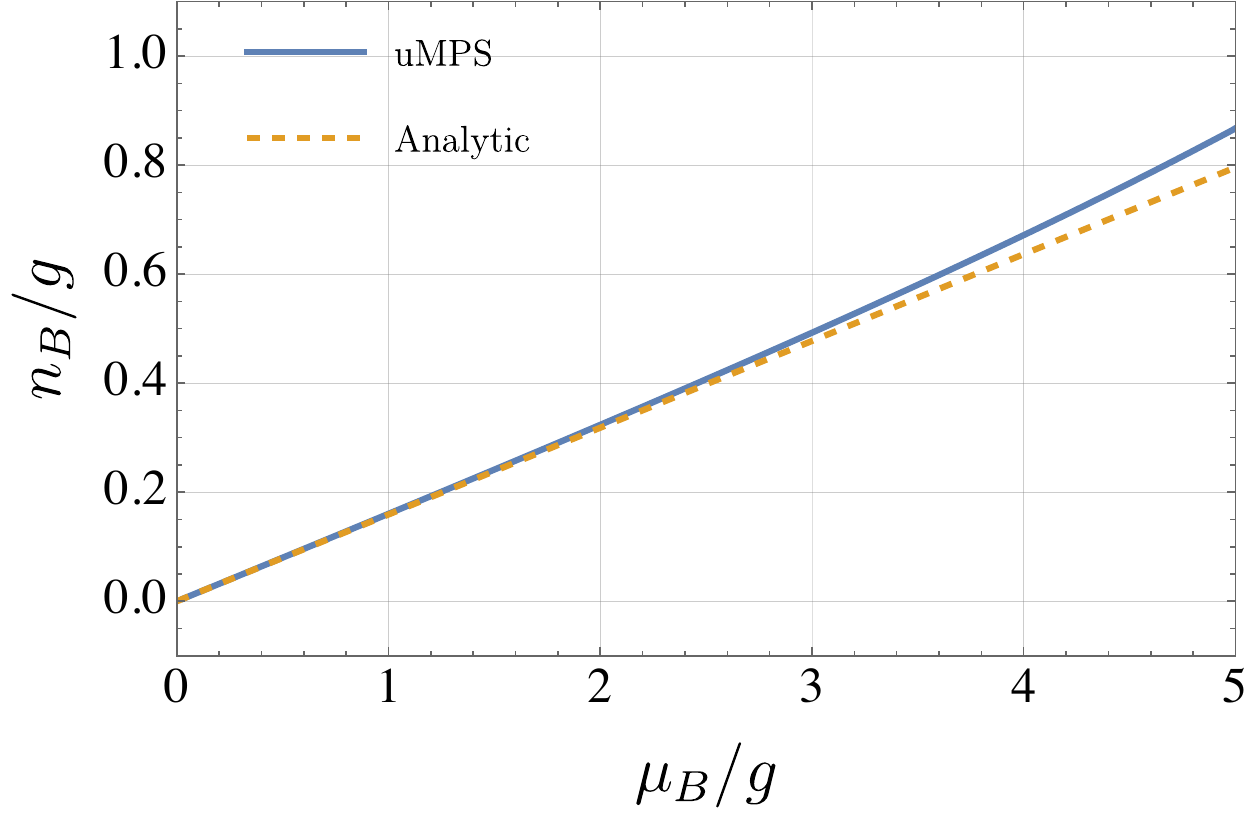}
        \end{minipage}
        \begin{minipage}[t]{0.45\textwidth}
        \centering
        \includegraphics[width=\textwidth]{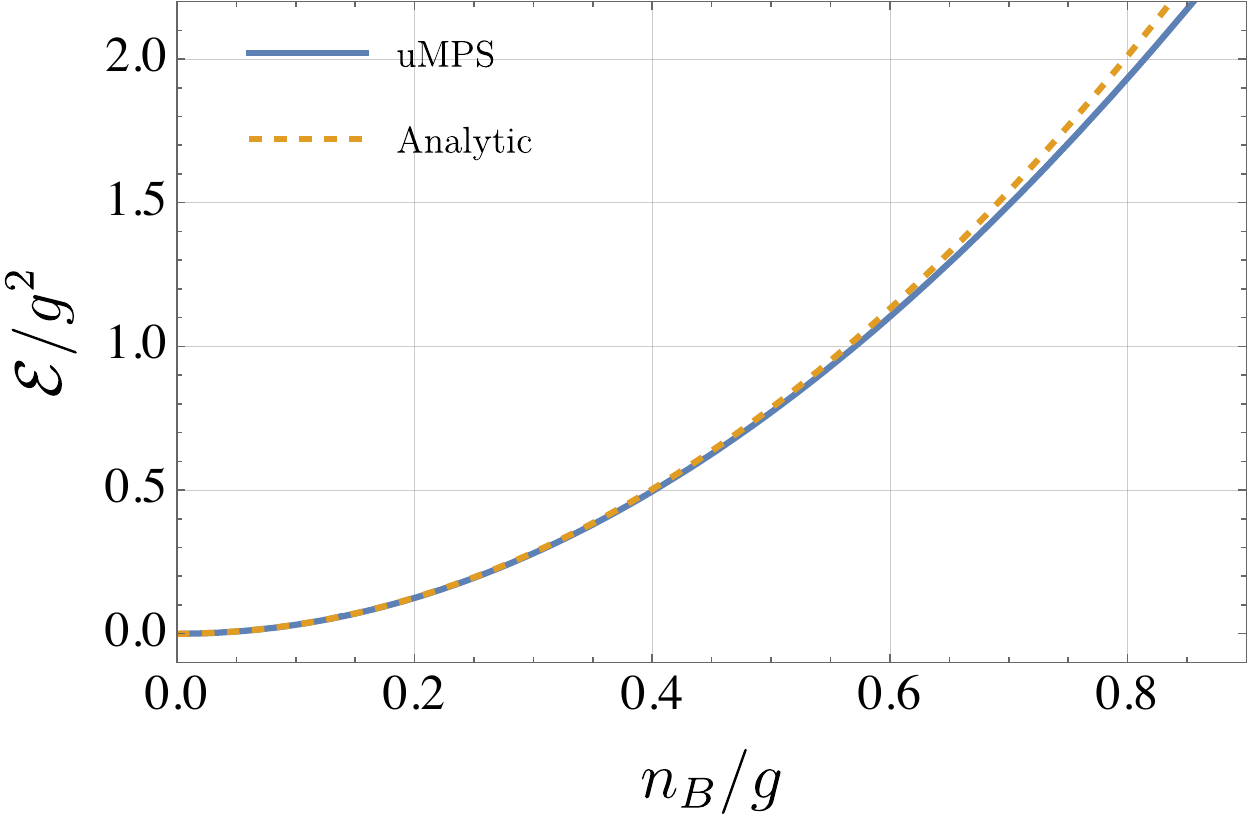}
        \end{minipage}
\caption{
Dependence of the baryon number density on the baryon chemical potential (left), and dependence of the energy density on the baryon number density (right) for zero quark mass, $m/g=0$.
The blue curve shows the numerical result at fixed lattice spacing, $ga=0.25$, while the black dotted line represents the exact continuum result obtained by the bosonization method, $n_B =\mu_B/(2\pi)$ and $\mathcal{E}=\pi n_B^2$.
}\label{fig:comparison with analytic and lattice result}
\end{figure*}
We first consider the massless quark case, $m/g = 0$, for which an analytic solution is available in the continuum limit via bosonization (see appendix \ref{appendix:bosonization} for details):
\begin{align}
P=\frac{\mu_B^2}{4\pi},\quad
n_B=\frac{\mu_B}{2\pi},\quad
\ed=\pi n_B^2.
\label{eq:EOS_m=0}
\end{align}
Figure \ref{fig:comparison with analytic and lattice result} shows the dependence of the baryon number and energy densities on the baryon chemical potential.
As seen from the figure, the numerical lattice results are in good agreement with the exact results for small baryon number density.
At larger densities, however, deviations from the exact result become visible.
These deviations are likely due to lattice artifacts.
In fact, the Fermi momentum is given by $k_F = \pi n_B $.
As $n_B$ increases, the typical momentum scale approaches the lattice cutoff, and finite-lattice-spacing effects therefore become more significant.
Nevertheless, even though the system is gapless for any $n_B$ at $m/g=0$, the numerical results remain in good overall agreement with the analytic continuum prediction.

\begin{figure*}[htbp]
    \centering
    \begin{minipage}[t]{0.45\textwidth}
        \centering
        \includegraphics[width=\textwidth]{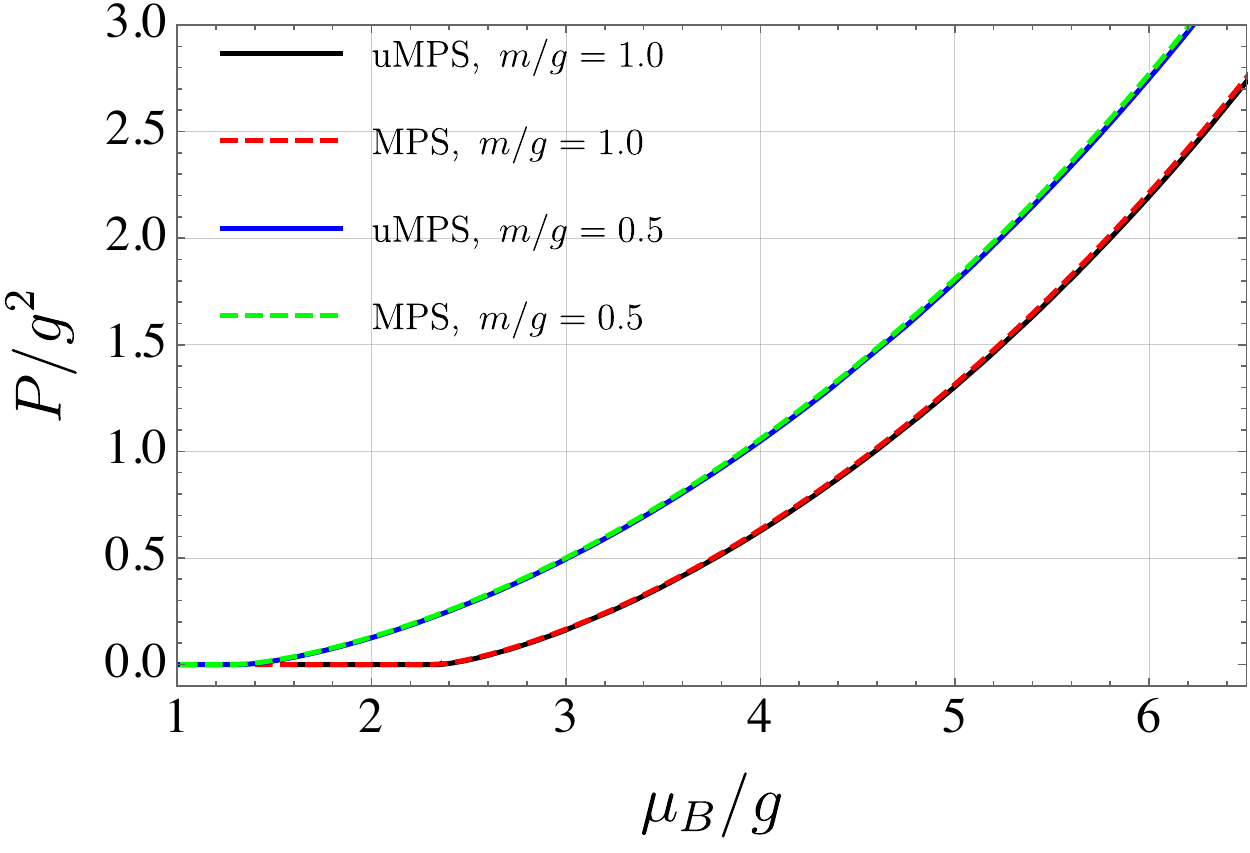}
    \end{minipage}
    \hspace{0.01\textwidth} 
    \begin{minipage}[t]{0.45\textwidth}
        \centering
        \includegraphics[width=\textwidth]{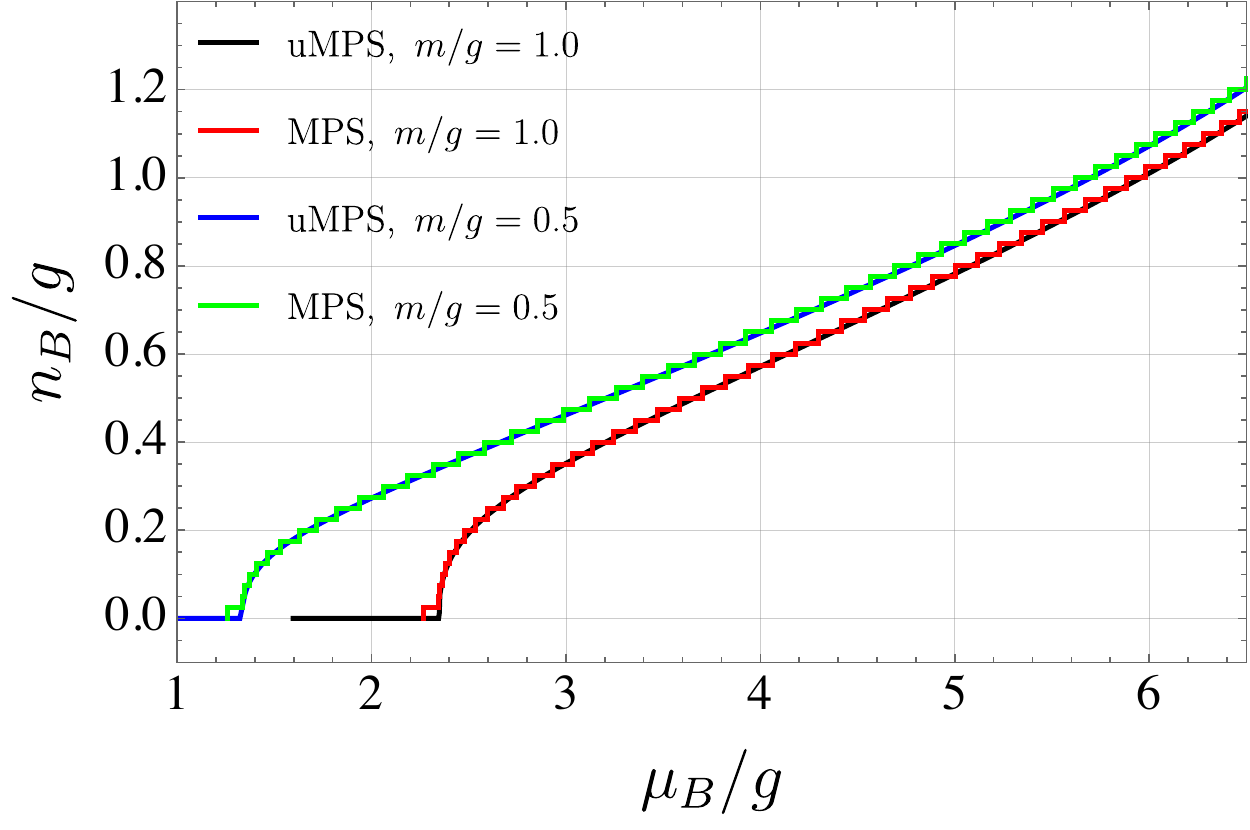}
    \end{minipage}
        \begin{minipage}[t]{0.45\textwidth}
        \centering
        \includegraphics[width=\textwidth]{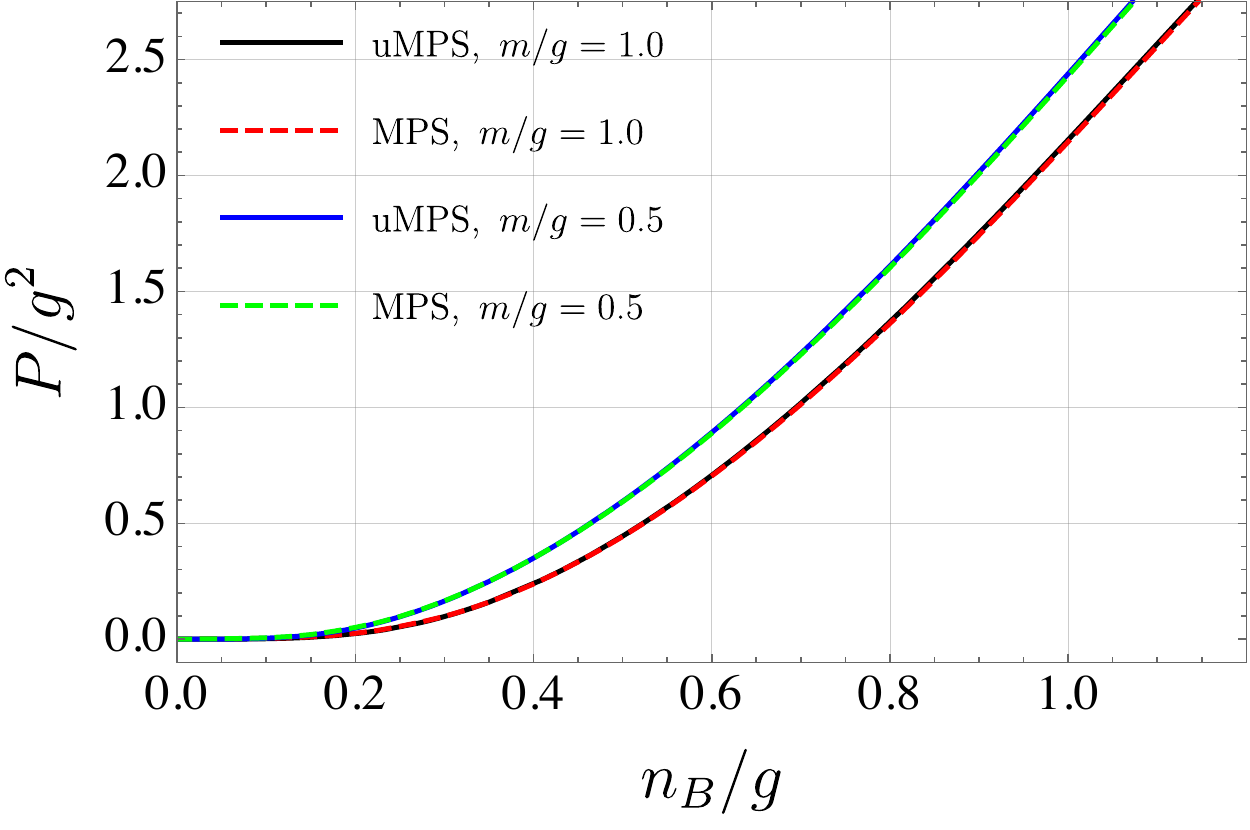}
    \end{minipage}
    \hspace{0.01\textwidth} 
    \begin{minipage}[t]{0.45\textwidth}
        \centering
        \includegraphics[width=\textwidth]{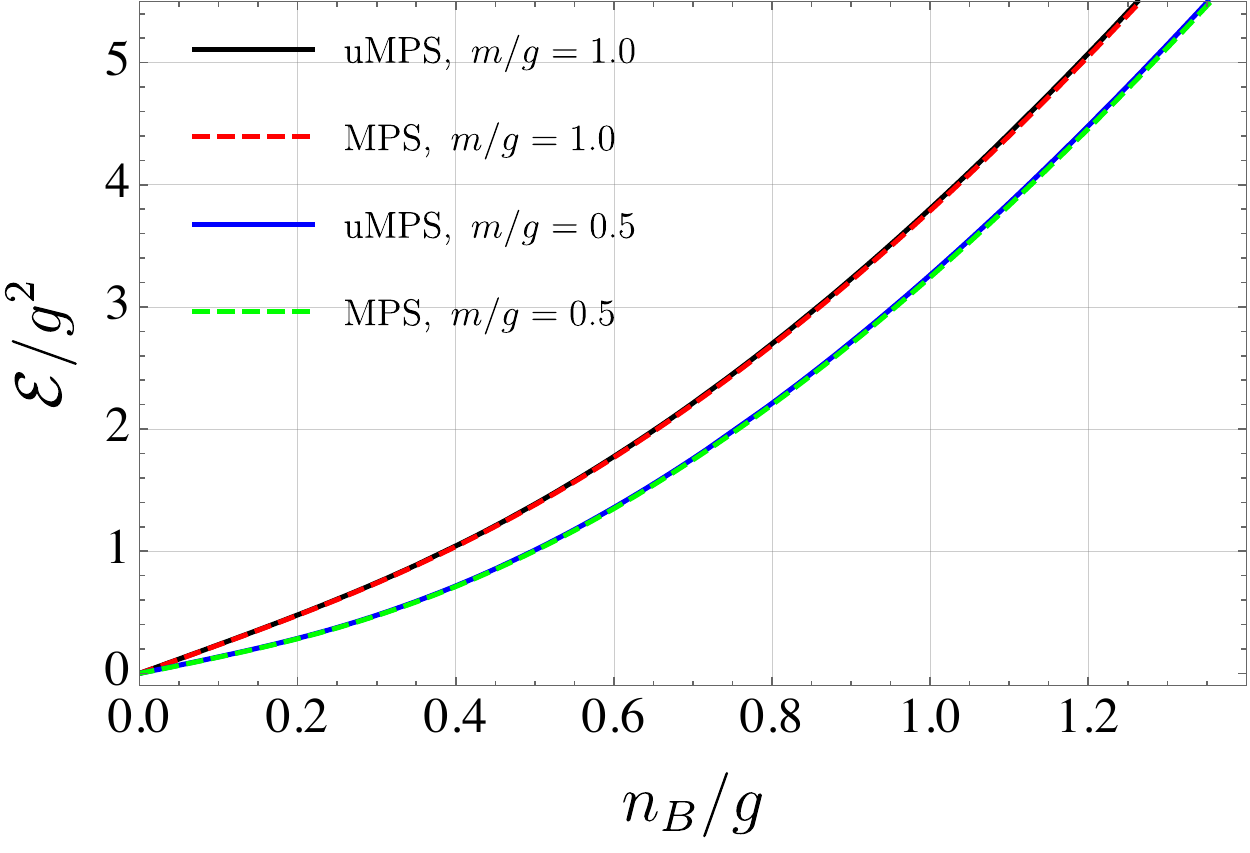}
    \end{minipage}
\caption{
Dependence of the pressure $P$ on the baryon chemical potential $\mu_B$ (top left), the baryon number density $n_B$ on $\mu_B$ (top right),  $P$ on $n_B$ (bottom left), and the energy density $\ed$ on $n_B$ (bottom right) for quark masses $m/g=0.5$ and $m/g=1.0$ at fixed lattice spacing $ga=0.25$.
The dashed and solid curves correspond to the finite-interval MPS results with open boundary conditions (ref.~\cite{Hayata:2023pkw}) and the uMPS results (the present work), respectively.
}\label{fig:thermodynamical quantities}
\end{figure*}
We next consider the case of nonzero quark mass, $m/g \neq 0$.
Figure~\ref{fig:thermodynamical quantities} shows various thermodynamic quantities together with results from the previous study based on DMRG calculations performed on a finite interval with open boundary conditions~\cite{Hayata:2023pkw}.
We find that the pressure, baryon number density, and energy density are in good agreement with those of the previous study.

A striking feature of the previous study is that the change in the baryon number density with respect to the baryon chemical potential is discrete due to finite-size effects.
In contrast, a smooth dependence on the baryon chemical potential is observed in the present study using uniform MPS, enabling a more accurate determination of thermodynamic quantities obtained by differentiation with respect to the baryon chemical potential.
This is one of the advantages of using the uniform MPS for simulations of the dense $\mathrm{QC_2D_2}$.

Another feature of the previous study is that an inhomogeneous phase was observed on a finite interval~\cite{Hayata:2023pkw}. 
In our calculation, translational invariance is imposed through the uMPS ansatz; therefore, such an inhomogeneous structure does not appear.
However, in (1+1) dimensions, continuous symmetries cannot be spontaneously broken in the thermodynamic limit~\cite{Coleman:1973ci,Mermin:1966fe,Hohenberg:1967zz}. 
The absence of spatial modulation in our results is consistent with this general theorem.
Moreover, we find that for $ n_B \neq 0$, the system is gapless and the two-point functions exhibit power-law decay, indicating enhanced long-range fluctuations. 
This behavior implies that the system is highly sensitive to small explicit breaking of translational invariance, such as boundary effects.
We therefore conclude that the inhomogeneous structure observed on a finite interval is likely induced by the explicit breaking term (i.e., boundary effects) and does not survive in the thermodynamic limit.

A similar consideration can be applied to the $\mathrm{U}(1)_B$ and $\mathrm{U}(1)_A$ symmetries,
where the latter corresponds to a translation by one site in the staggered fermion formulation.
In $(3+1)$ dimensions, a superfluid state, where $\mathrm{U}(1)_B$ symmetry is spontaneously broken by the diquark condensate, appears at a certain baryon chemical potential, as confirmed by the lattice studies.\footnote{The conventional lattice Monte-Carlo simulation can be applied to two-color QCD with an even number of flavors because there is no sign problem (See, e.g., ref.~\cite{Itou:2025vcy} for a review.)}
However, such a phase is also disfavored in $(1+1)$ dimensions because continuous symmetries cannot be spontaneously broken.
We numerically confirm that the expectation value of the diquark operator vanishes in the limit of large bond dimensions.
On the other hand, the baryon number density operator can develop a nonzero condensate.

We discuss the physical properties near the onset of nonzero baryon number density.
As mentioned in the previous subsection, the excitations form a triplet.
At finite density, the chemical potential explicitly breaks the $\mathrm{SU}(2)/\mathbb{Z}_2$ symmetry down to $\mathrm{U}(1)_B$, so that the degeneracy is lifted.
The only baryon with unit baryon number becomes lighter, and for $\mu_B > M_B$ a finite baryon density is expected to develop.
In the dilute baryon-gas region, \(0 \leq n_B/g \leq 0.05\), with the lattice spacing fixed at \(ga=0.25\), the VUMPS algorithm does not converge and cannot construct the optimized ground state.
We monitor the optimization process and find that the VUMPS algorithm cannot distinguish between the vacuum state and a dilute finite-density state.
Although we increase the accuracy of the iterative solvers implemented in the VUMPS algorithm to resolve these states, no practical improvement is observed.
The degeneracy of the pressures of the vacuum and dilute finite-density states can be seen in Fig.~\ref{fig:thermodynamical quantities}.
In fact, the grand potential difference between a dilute finite-density state and the vacuum is proportional to $n_B^3$ when the ideal-gas and nonrelativistic approximations are valid for baryons (See appendix~\ref{appendix:bosonization} for discussions on the transition from vacuum state to the finite baryon density state using bosonization method).

\begin{figure*}[htbp]
    \centering
    \begin{minipage}[t]{0.45\textwidth}
        \centering
        \includegraphics[width=\textwidth]{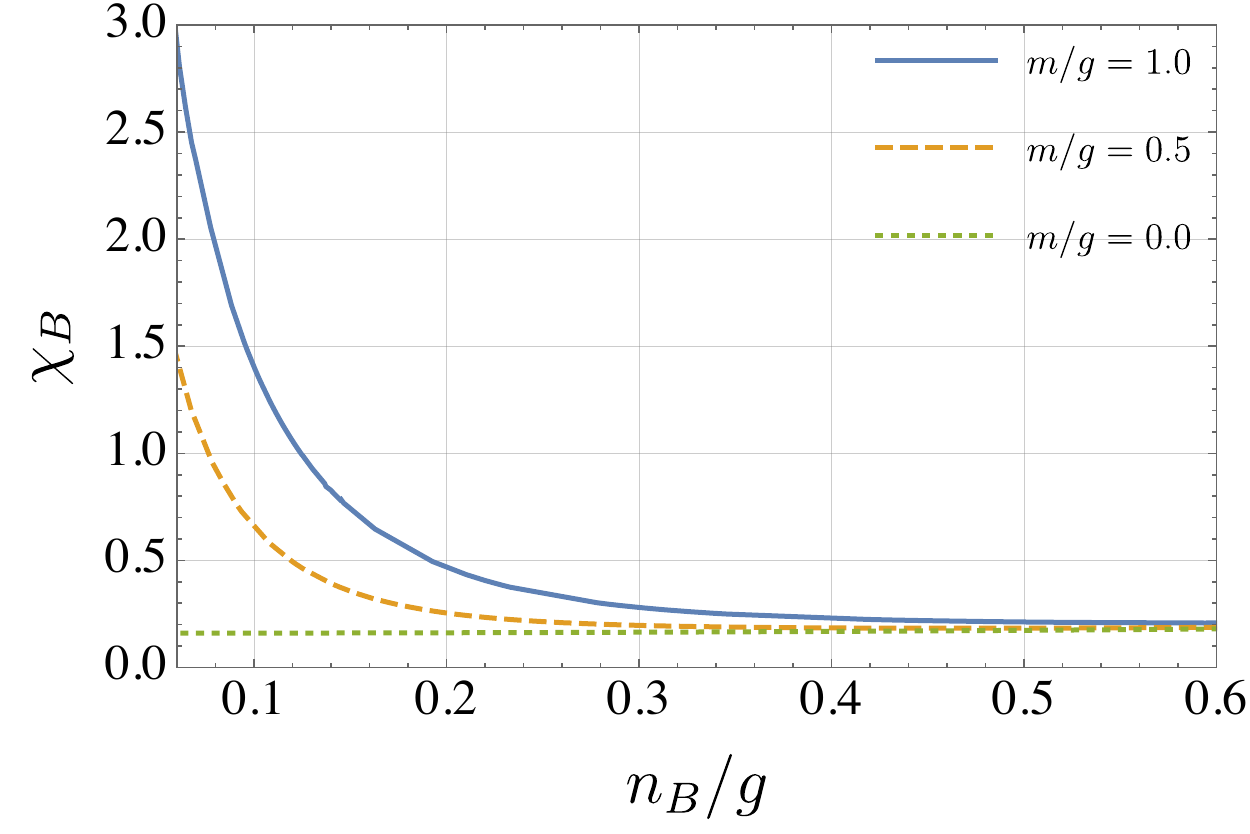}
    \end{minipage}
\caption{
The baryon-number susceptibilities $\chi_B$ for quark masses $m/g=1.0$, $m/g=0.5$ and $m/g=0.0$ are shown at a fixed lattice spacing, $ga=0.25$.
}\label{fig:baryon number susceptibility}
\end{figure*}
To further characterize the onset of nonzero baryon number density, we evaluate the baryon-number susceptibility at zero-temperature, defined by 
\begin{align}
    \chi_B =\dfrac{\mathrm{d}^2 P}{\mathrm{d}\mu_B^2}=\dfrac{\mathrm{d}n_B}{\mathrm{d}\mu_B}.
\end{align}
Figure~\ref{fig:baryon number susceptibility} shows the dependence of \(\chi_B\) on \(n_B\) for several quark masses.
As shown in the figure, for \(m/g \neq 0\), \(\chi_B\) grows sharply and is consistent with a divergence as the onset of nonzero baryon number density is approached, \(n_B \to 0\).
This behavior supports the scenario of a second-order phase transition at the onset of nonzero baryon number density.
However, the possibility of a weak first-order phase transition with a jump in the order parameter within the range \(0 \leq n_B/g \leq 0.05\), or of a sharp crossover, cannot be excluded.
$m/g=0$ is an exactly solvable limit leading to $\chi_B=1/(2\pi)$, which does not depend on $n_B$.

\subsubsection{Low-Energy Behavior as a Tomonaga--Luttinger Liquid}

\begin{figure*}[htbp]
    \centering
    \begin{minipage}[t]{0.45\textwidth}
        \centering
        \includegraphics[width=\textwidth]{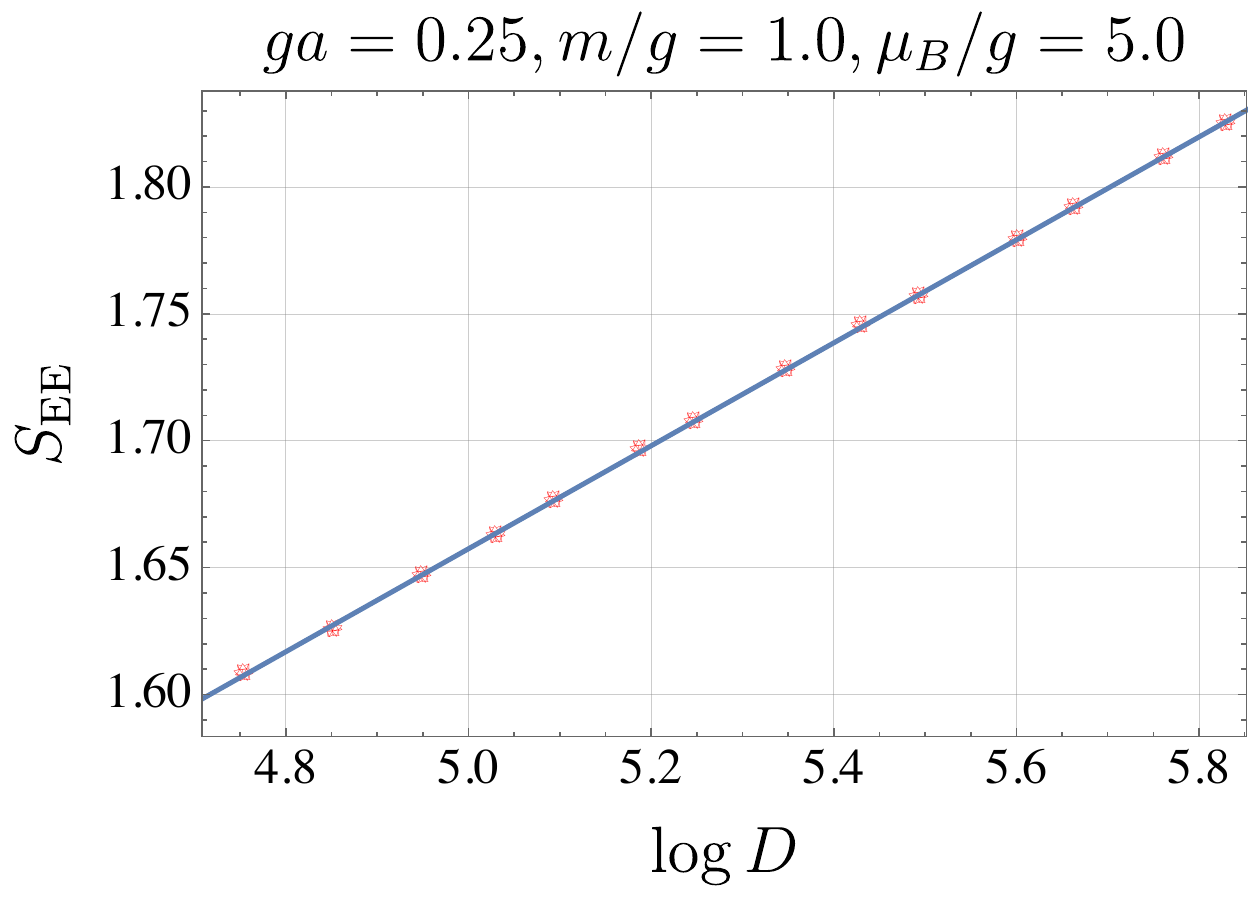}
    \end{minipage}
    \hspace{0.01\textwidth} 
    \begin{minipage}[t]{0.45\textwidth}
        \centering
        \includegraphics[width=\textwidth]{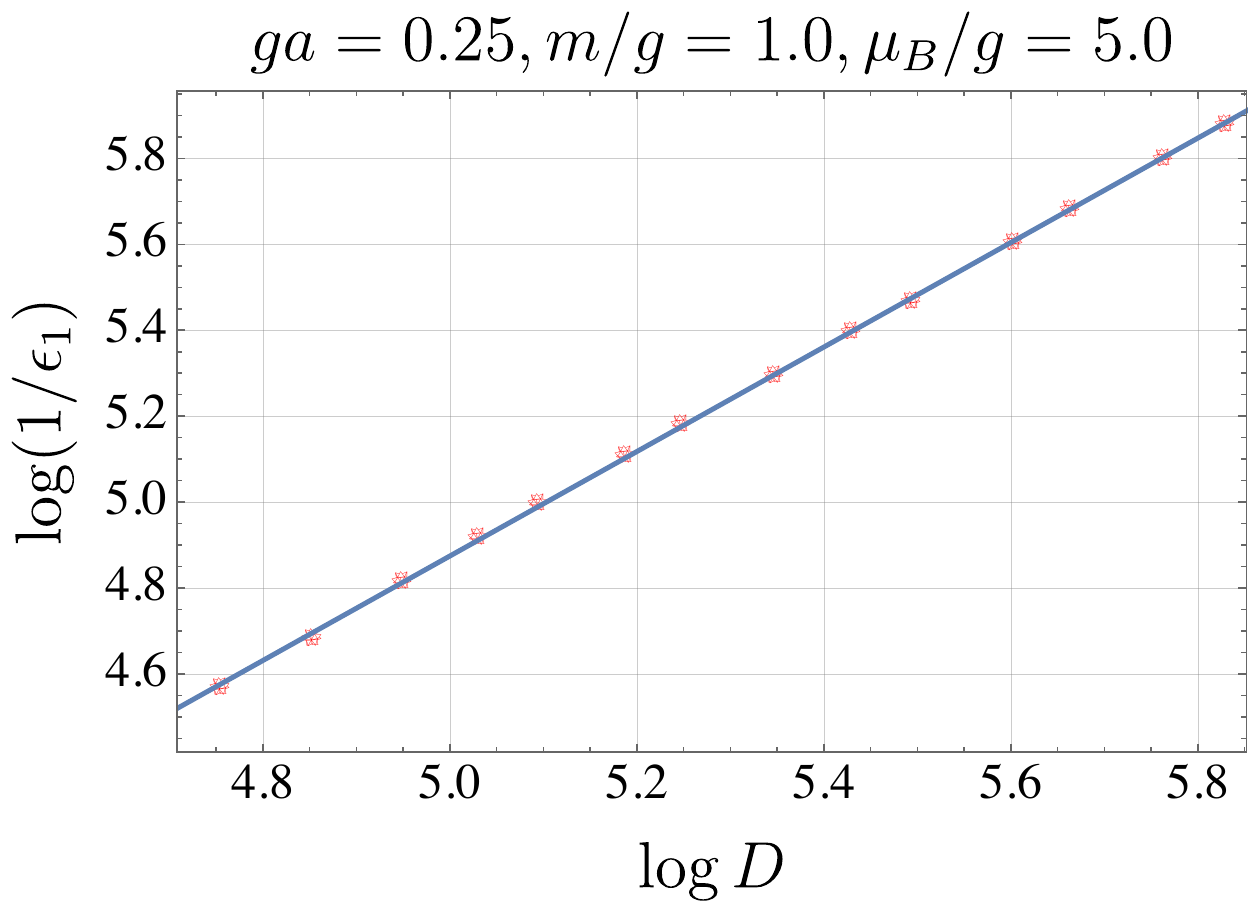}
    \end{minipage}
\caption{
Dependence of the entanglement entropy $S_{\rm EE}$ (left) and the correlation length
$\xi=1/\epsilon_1$ in lattice units (right) on the total bond dimension $D$ at finite density ($ga=0.25,~m/g=1,~\mu_B/g=5$).
Red symbols and blue lines represent the numerical data and linear fits, respectively.
}\label{fig:entanglement scaling at finite density}
\end{figure*}
In this section, we show that the low-energy behavior of the system at finite baryon density is consistent with a Tomonaga--Luttinger liquid.
To this end, we further discuss the ground-state properties using transfer matrices.
For a large baryon chemical potential where $ n_B \neq 0$, we find that the correlation length diverges as we decrease the truncation error (and hence increase the bond dimension).
Figure \ref{fig:entanglement scaling at finite density} shows the entanglement entropy between two semi-infinite subsystems separated at a given point, and the correlation length as a function of the bond dimension for fixed lattice input parameters $ga=0.25,~m/g=1,~\mu_B/g=5$.
Even for $m/g\neq 0$, the system exhibits the scaling behavior $\log(1/\epsilon_1)\propto \log(D)$ with remarkable accuracy, which indicates a gapless phase.
Motivated by this observation, we estimate the central charge by fitting $S_\mathrm{EE}\propto \log D$, as described in the previous subsection.
We find that the central charge is $c=1.00$ within the fitting error.
We also compute the central charge for the free fermions by setting the gauge coupling to zero, and we obtain $c=2$.
This indicates that the gauge interaction remains effective in the long-distance limit in the presence of the finite baryon density.
Note that $c=1$ coincides with that of the free compact boson, which is identified as the baryon number density mode in $\mathrm{SU}(2)$ gauge theory.
For different values of the chemical potential, we observe the same behavior as long as the baryon number density is nonvanishing.
This result has a natural interpretation in the quarkyonic picture~\cite{McLerran:2007qj,Kojo:2009ha,Kojo:2011cn}: the reduction from $c=2$ for free quarks to $c=1$ reflects the fact that color confinement gaps out quark-like excitations, leaving the baryon density mode as the only gapless degree of freedom in the infrared.

The infrared behavior at $n_B \neq 0$ is argued to be described by the Tomonaga--Luttinger liquid theory via bosonization~\cite{Lajer:2021kcz}.
We briefly review the bosonization of $\mathrm{QC_2D_2}$, and Tomonaga--Luttinger liquid theory in appendix~\ref{appendix:bosonization}.
The effective Hamiltonian in the long-distance limit is given by
\begin{align}
    H_\mathrm{TLL} (\mu_B>M_B) =\int\mathrm{d}x\,\left[2\pi c_sK \Pi_{B}^2+\dfrac{c_s}{8\pi K}(\partial_1\tilde{\phi}_B)^2\right].
\end{align}
Here, $\tilde{\phi}_B$ is the real compact scalar with period $2\pi$, while $\Pi_B$ is its canonical conjugate.
The parameters $c_s$ and $K$ denote the speed of sound and the Luttinger parameter, respectively.
The speed of sound at zero temperature is estimated by the equation of state,
\begin{align}
    c_s^2=\dfrac{\mathrm{d}P}{\mathrm{d}\ed}=\frac{\mathrm{d}P}{\mathrm{d}n_B}\frac{\mathrm{d}n_B}{\mathrm{d}\ed}=\frac{n_B}{\mu_B}\frac{\mathrm{d}\mu_B}{\mathrm{d}n_B}.
\end{align}
The Luttinger parameter is defined by the relation [see eqs.~\eqref{eq:Luttinger1}-\eqref{eq:Luttinger4}],
\begin{align}
    K= \frac{\pi}{c_s} \frac{n_B}{\mu_B}.
\end{align}
For $m/g=0$, eq.~\eqref{eq:EOS_m=0} yields $P=\ed$ and $n_B=\mu_B/(2\pi)$ for $N_c=2$.
These relations imply $c^2_s=1$, and $K=1/2$.
For general $\mu_B$ with $m/g \neq 0$, there is no simple relation between $(K,c_s)$ and $(m/g,\mu_B/g)$.

\begin{figure*}[htbp]
    \centering
    \begin{minipage}[t]{0.45\textwidth}
        \centering
        \includegraphics[width=\textwidth]{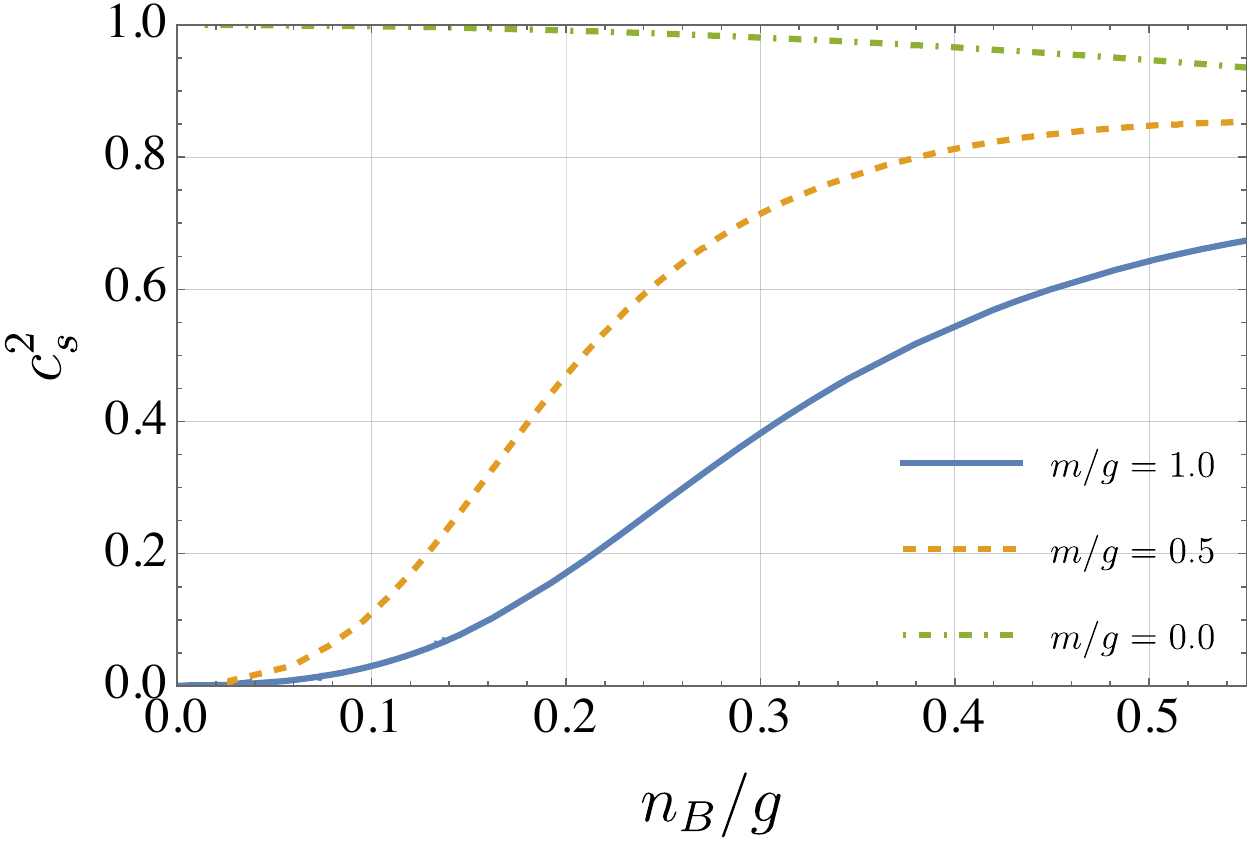}
    \end{minipage}
    \hspace{0.01\textwidth} 
    \begin{minipage}[t]{0.45\textwidth}
        \centering
        \includegraphics[width=\textwidth]{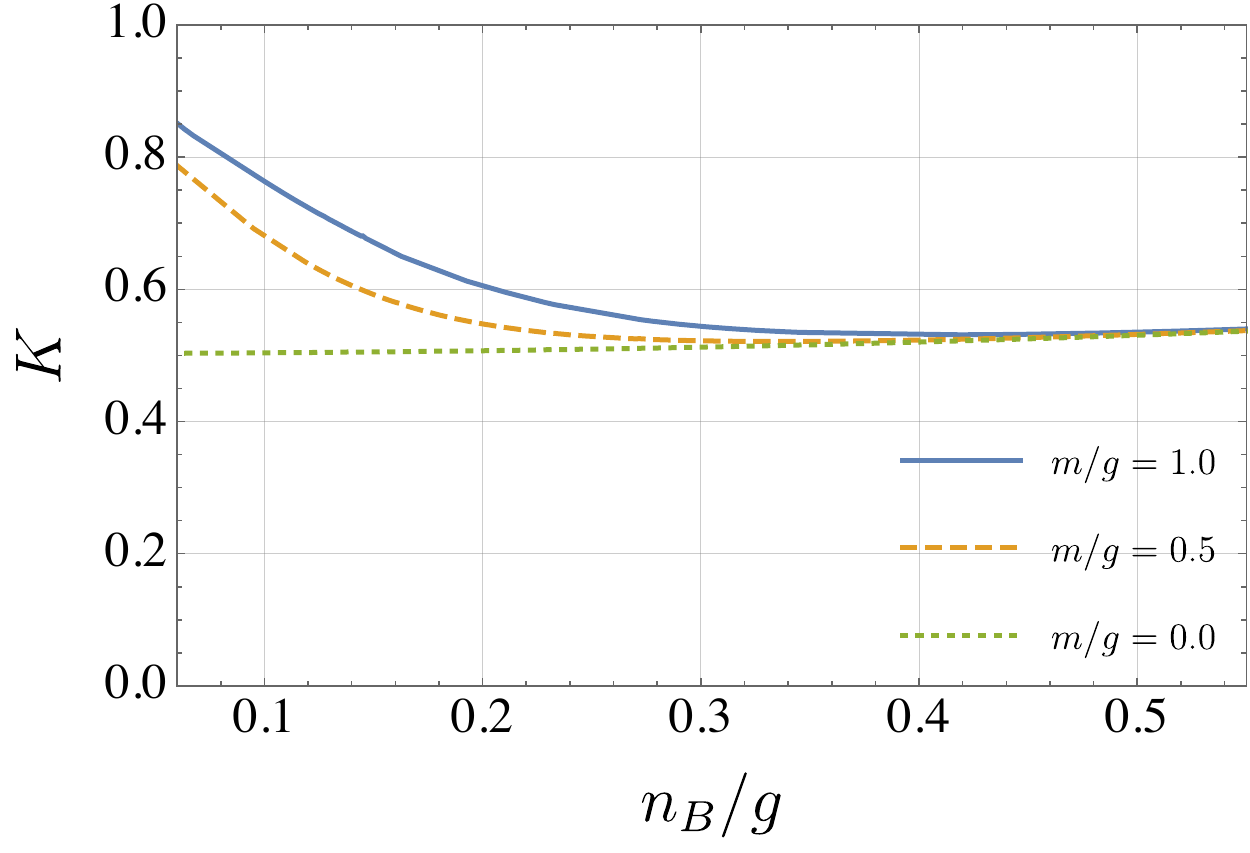}
    \end{minipage}
\caption{Speed of sound squared $c_s^2$ extracted from the equation of state (left) and the Luttinger parameter $K$ (right), shown as functions of the baryon number density for different quark masses.
}\label{fig:sound speed and Luttinger parameter}
\end{figure*}
Figure \ref{fig:sound speed and Luttinger parameter} shows the dependence of the speed of sound and the Luttinger parameter on the baryon number density.
At $m/g=0$, we find agreement with the analytic result for both $c_s=1$ and $K=1/2$.
The small deviations in the high-density region can be understood as lattice artifacts.
For a nonvanishing quark mass $m/g\neq 0$, the overall behavior changes drastically.
In particular, $c_s\to 0$ is found at the transition point $ n_B\to 0$, as we expected.
Unfortunately, we cannot reliably determine the Luttinger parameter at this transition point, since both $\mathrm{d}\mu_B/\mathrm{d}n_B\to 0$ and $ n_B\to 0$, which make numerical computations unstable.
The density dependence of the Luttinger parameter $K$ provides a quantitative characterization of the baryon--quark crossover.
At low density, $K \simeq 1$ indicates a baryonic ideal-gas regime, where baryons, identified with kinks, behave as free fermions, or equivalently as hard-core bosons, with compactification radius $R^2 = 1/(4\pi)$.
At high density, by contrast, $K \to 1/N_c$, reflecting the quark-like nature of the dense matter, in which the baryon number is carried by $N_c$ quark degrees of freedom.
The smooth interpolation of $K(n_B)$ from $1$ to $1/N_c$ therefore traces the continuous crossover from baryonic matter to quarkyonic matter, consistent with the analytical results in Appendix~\ref{appendix:bosonization}.

When the system is described by the Tomonaga--Luttinger liquid, the equal-time two-point function of the baryon number density operator $j^0_B$ takes the form~\cite{Haldane:1981zz},
\begin{align}
    \langle j^0_B(t,x_1)j^0_B(t,x_2)\rangle-n_B^2=\dfrac{B_0}{|x_1-x_2|^2} +\sum_{\ell=1}^{\infty} B_\ell\dfrac{\cos(2\pi \ell n_B |x_1-x_2|)}{|x_1-x_2|^{2\ell^2 K}},
\end{align}
where $B_\ell$ are constants. This expression implies that the two-point function exhibits spatial modulation.
Comparing this expression with the corresponding lattice expression eq.~\eqref{eq:lattice version of two-point function}, we expect to observe complex phases in the eigenvalues of the transfer matrix, 
$\phi_{k_\ell}=4\pi \ell a n_B$ ($\ell=1,2,\cdots$) corresponding to the physical momentum $p_{\ell}=2\pi \ell n_B$.

\begin{figure*}[htbp]
    \centering
    \begin{minipage}[t]{0.45\textwidth}
        \centering
        \includegraphics[width=\textwidth]{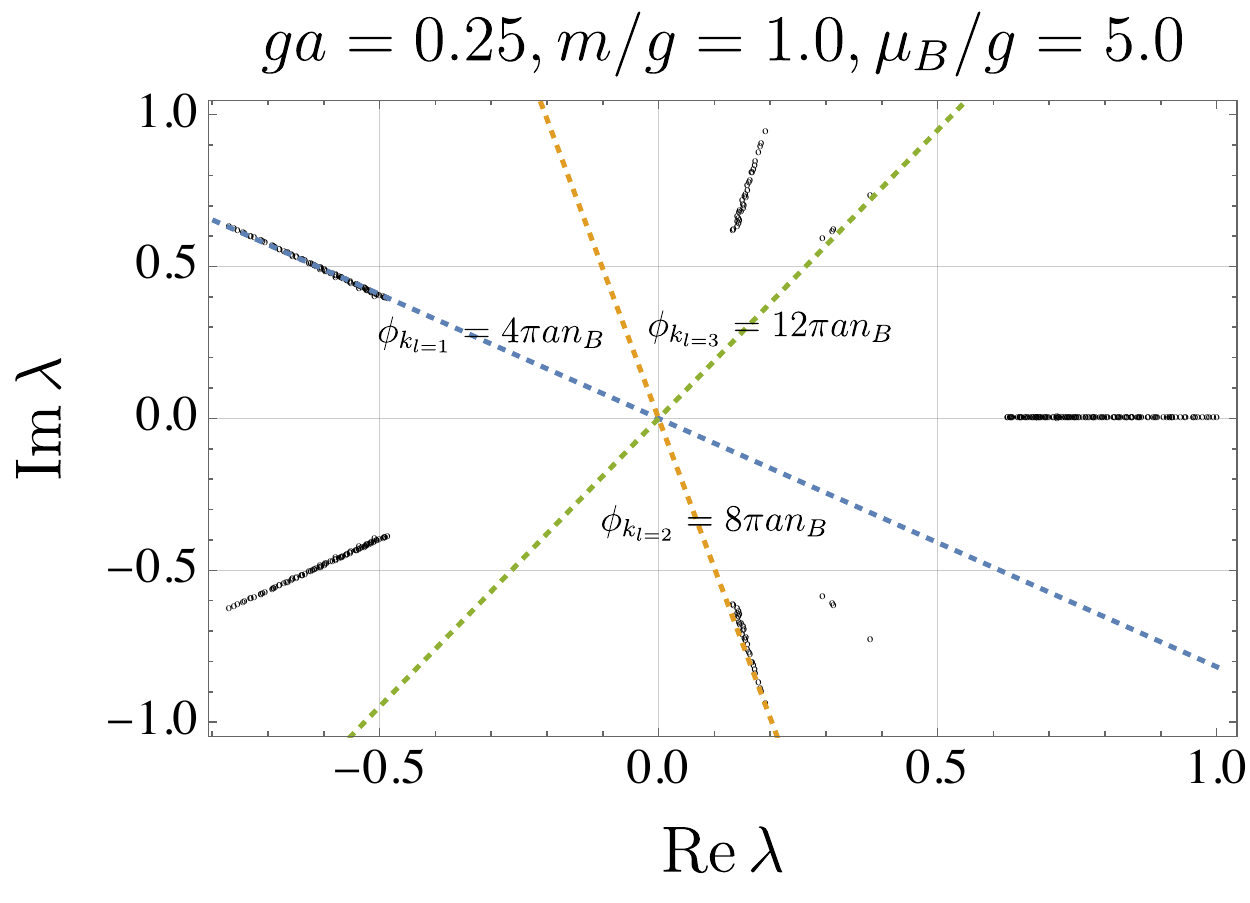}
    \end{minipage}
    \hspace{0.01\textwidth} 
    \begin{minipage}[t]{0.45\textwidth}
        \centering
        \includegraphics[width=\textwidth]{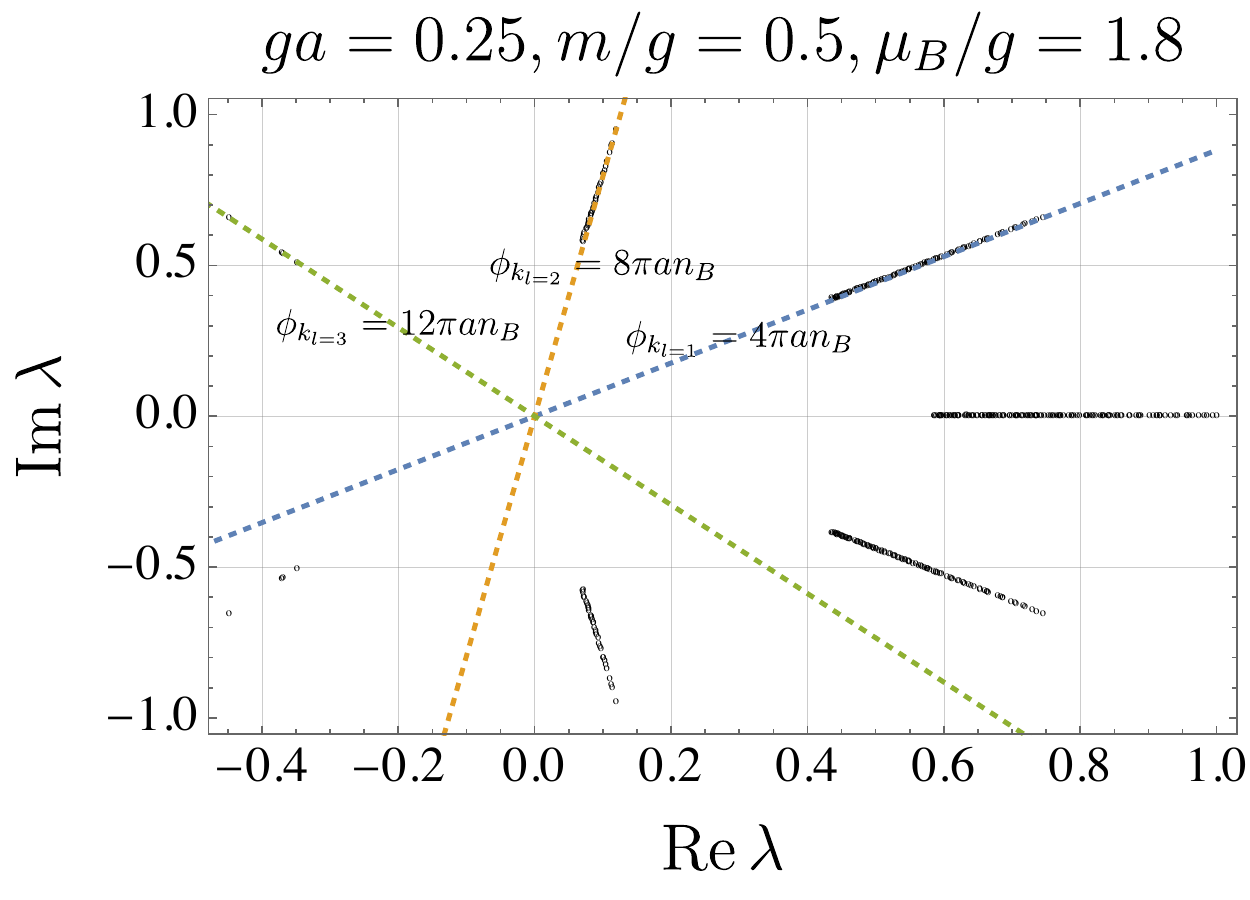}
    \end{minipage}
\caption{
Eigenvalues of the transfer matrix $\lambda$ on the complex plane for $m/g=1.0$, $\mu_B/g=5.0$ (left) and $m/g=0.5$, $\mu_B/g=1.8$ (right). 
Dotted lines indicate modulations with wavenumber $\phi_{k_\ell} = 4\pi \ell an_B$, where $\ell = 1$ (blue), $\ell = 2$ (orange), and $\ell = 3$ (green). Here, $n_B$ denotes the expectation value of the baryon number density.
}\label{fig:distribution of eigenvalues of TM}
\end{figure*}

Figure \ref{fig:distribution of eigenvalues of TM} shows the distribution of eigenvalues of the transfer matrix $T_{\mathcal{C}}$ up to $k\leq 400$.
In the figure, the lines corresponding to complex phases $\phi_{k_{\ell}}$ are also indicated.
The complex phases corresponding to the $\ell=1$ and $\ell=2$ modes are clearly visible, whereas the $\ell=3$ mode appears only marginally.
Contributions from higher modes ($\ell\geq 4$) cannot be resolved, likely due to finite lattice spacing effects or the truncation associated with a finite bond dimension.
Therefore, these results indicate that the low-energy behavior of the system is consistent with a Tomonaga--Luttinger liquid.
We confirm that the same behavior is observed for different values of the chemical potential as long as the baryon number density is nonvanishing.

\subsubsection{Quark distribution functions}
To probe the quark-level structure of the finite-density ground state and provide direct evidence for the quarkyonic picture, we compute the quark distribution function, which characterizes how quarks populate momentum states in the ground state.
The quark distribution function in momentum space can be expressed as
\begin{align}
    n(p)=\frac{1}{N_c}D(p)-1,~D(p)=\sum_{n=-\infty}^{\infty}e^{ip(2an)}W(-n/2,n/2).
\end{align}
Here, $D(p)$ is defined as the Fourier transform of the quark two-point function in position space $W(n_1,n_2)$ which is given by
\begin{align}
    W(n_1,n_2)=(-1)^{n_1-n_2}\left\langle \phi^{\dag}(2n_1)\left(\prod^{2n_1-1}_{s=2n_2}U(s)\right)\phi(2n_2)+\phi^{\dag}(2n_1+1)\left(\prod^{2n_1}_{s=2n_2+1}U(s)\right)\phi(2n_2+1)\right\rangle
\end{align}
for $n_1\geq n_2$ and 
\begin{align}
    W(n_1,n_2)=(-1)^{n_1-n_2}\left\langle \phi^{\dag}(2n_1)\left(\prod^{2n_2-1}_{s=2n_1}U(s)\right)^\dag\phi(2n_2)+\phi^{\dag}(2n_1+1)\left(\prod^{2n_2}_{s=2n_1+1}U(s)\right)^\dag\phi(2n_2+1)\right\rangle
\end{align}
for $n_1<n_2$.
The Wilson line insertion ensures gauge invariance of the two-point function.
Here, $p$ is the physical momentum.
In the limit $g\to0$, the Hamiltonian~\eqref{eq:Hamiltonian SU(2)} reduces to the free theory of two massive Dirac fermions, which is exactly solvable on the lattice (See appendix A of ref.~\cite{Hayata:2023pkw}).
In this case, $n(p)=\Theta(k_F-p)$ in the continuum and thermodynamic limits, where $\Theta(x)$ is the Heaviside step function, and $k_F =\pi n_B$ is the Fermi momentum.
We also construct the ground state of the free theory by VUMPS algorithm with $g=0$.

\begin{figure*}[htbp]
    \centering
        \begin{minipage}[t]{0.45\textwidth}
        \centering
        \includegraphics[width=\textwidth]{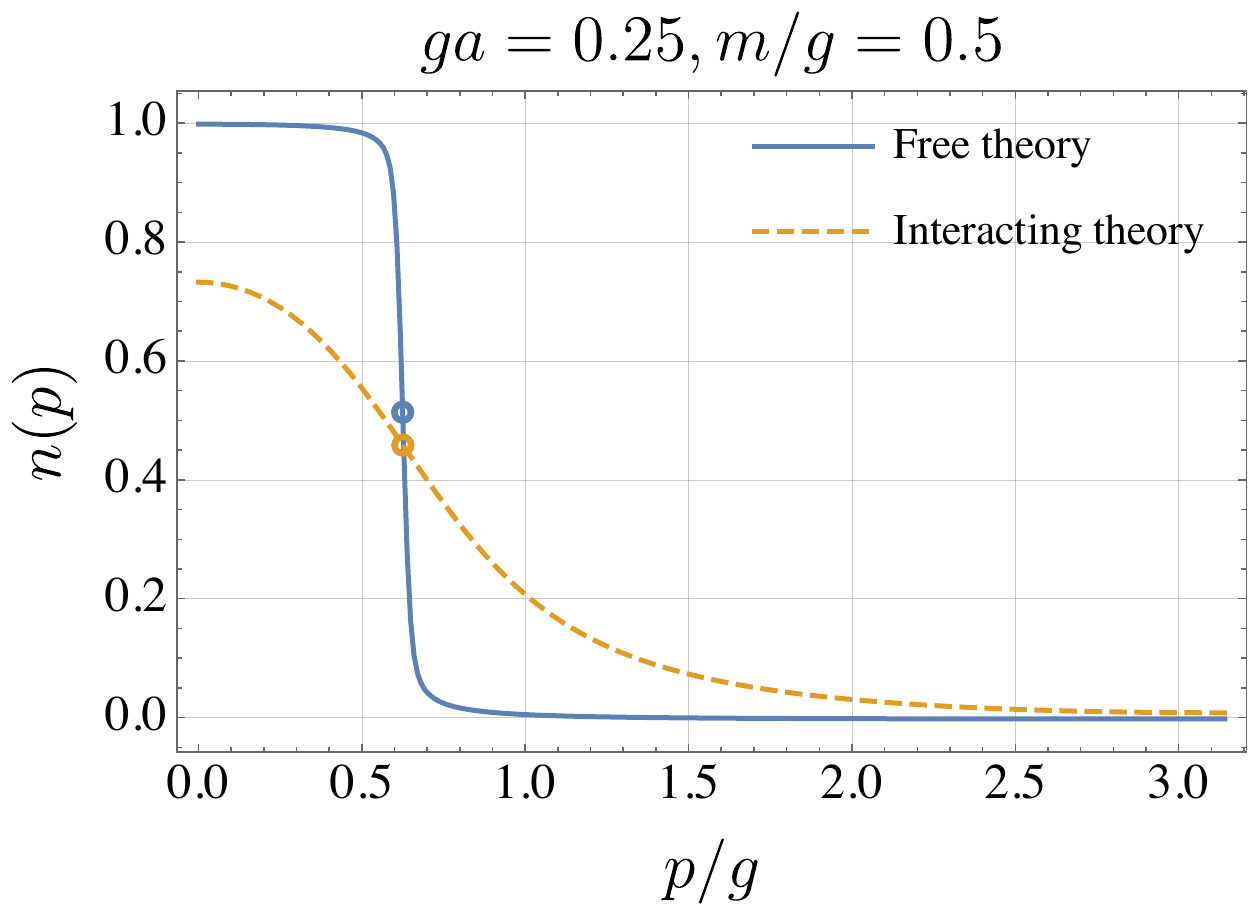}
    \end{minipage}
    \hspace{0.01\textwidth} 
    \begin{minipage}[t]{0.45\textwidth}
        \centering
        \includegraphics[width=\textwidth]{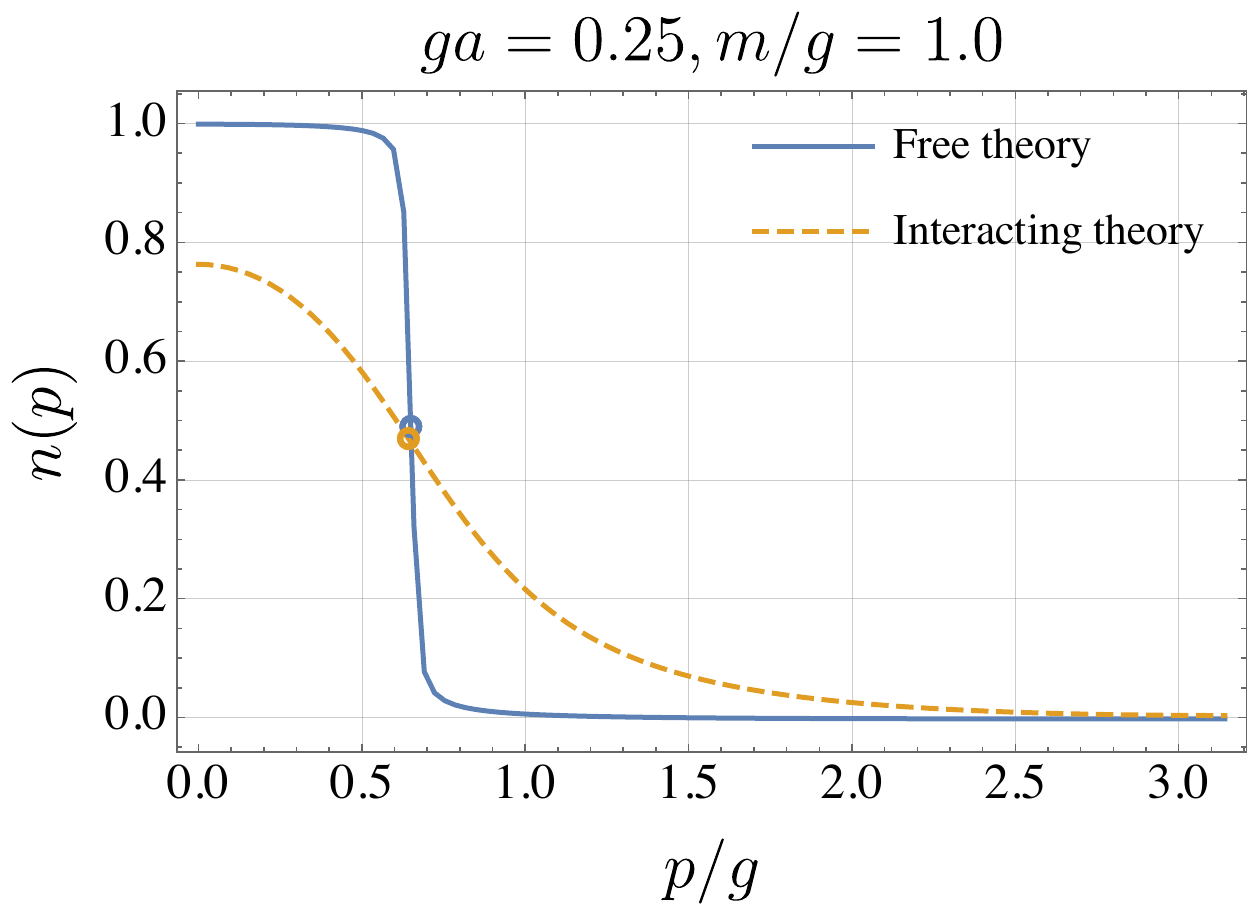}
    \end{minipage}
    \begin{minipage}[t]{0.45\textwidth}
        \centering
        \includegraphics[width=\textwidth]{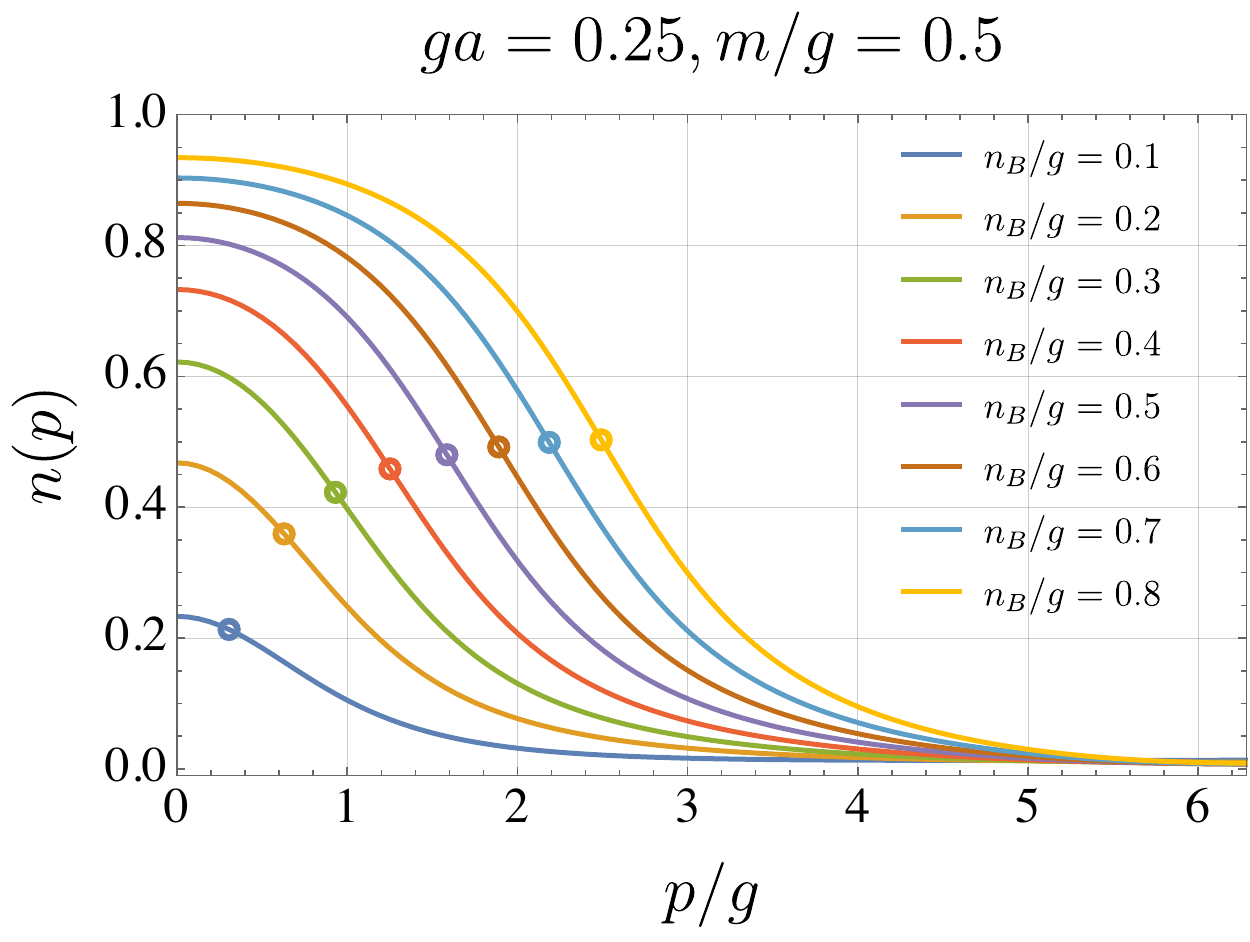}
    \end{minipage}
    \hspace{0.01\textwidth} 
    \begin{minipage}[t]{0.45\textwidth}
        \centering
        \includegraphics[width=\textwidth]{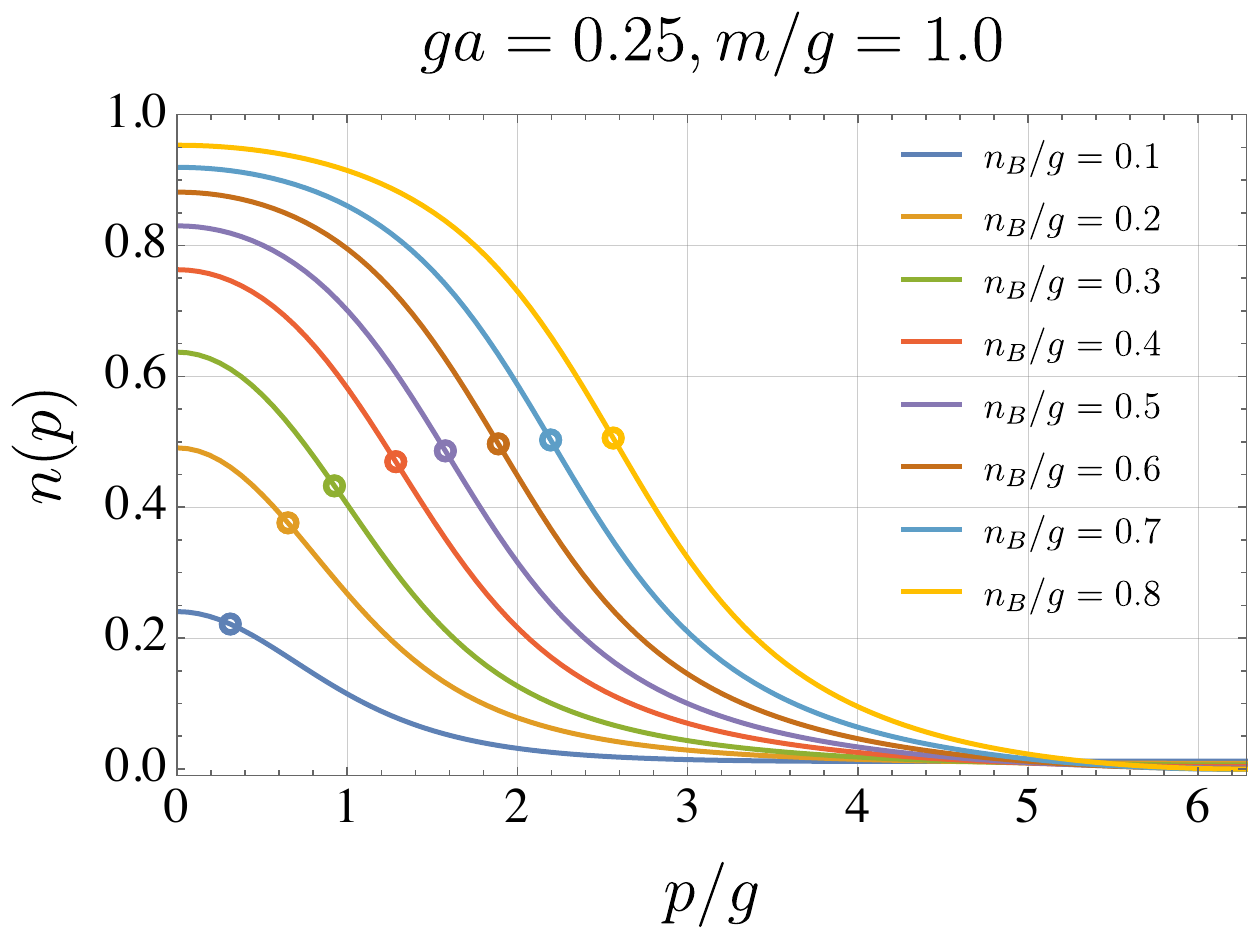}
    \end{minipage}
\caption{
The solid and dashed curves correspond to the cases without and with gauge interactions for the same baryon chemical potential $\mu_B/g=2.65$ (top left) and $\mu_B/g=3.24$ (top right), respectively.
Quark distribution functions in momentum space for $m/g = 0.5$ (bottom left) and $m/g = 1.0$ (bottom right) as functions of momentum for various values of the baryon number densities.
Circle points represent the Fermi momenta, $k_F = \pi n_B$.
}
\label{fig:quark distribution function}
\end{figure*}
The case without gauge interactions is obtained by omitting the electric-field term from the Hamiltonian. 
In one spatial dimension, the remaining link variables carry no energetic cost and can be gauged away, so that the model is equivalent to free staggered fermions.
The step-function behavior in the free theory is clearly observed.
Once the gauge interaction is included, the Fermi surface $p=k_F$ becomes unstable.
Remarkably, a depression of the quark number distribution at low momenta found in the previous study~\cite{Hayata:2023pkw} does not appear in our case, thanks to translational invariance.
Quark distribution functions for various baryon number densities with gauge interaction are also shown in the bottom panels of Fig.~\ref{fig:quark distribution function}.
Although there is no sharp Fermi surface, the quark number density at low momenta $p\leq k_F$ remains occupied as can be seen from the figure.
Therefore, the free quark picture provides a useful approximation for thermodynamic quantities, such as the baryon number density and pressure, in the high-density region, even in the presence of the gauge interaction.
However, we emphasize that the free quark picture cannot be applied to other physical quantities, such as the central charge.

It is worth noting that the behavior of the distribution function differs from that in electronic systems described by a Tomonaga--Luttinger liquid.
In electronic systems, the distribution function exhibits a power-law behavior near the Fermi surface~\cite{Voit:1995pfa,Giamarchi:2003}, whereas in our case it appears to decay more rapidly.
This difference is likely a consequence of color confinement in the present system.
The coexistence of a quark Fermi sea, characterized by $n(p) \approx 1$ for $p \lesssim k_F$, and a baryonic infrared description, namely a Tomonaga--Luttinger liquid with $c=1$, provides a concrete realization of the quarkyonic picture~\cite{McLerran:2007qj,Kojo:2011fh} in $(1+1)$-dimensional $\mathrm{SU}(2)$ gauge theory.
In this picture, the bulk properties of the Fermi sea are described in terms of quark-like degrees of freedom, while the low-energy excitations near the Fermi surface are baryonic in nature.

\section{Conclusion and Discussion}\label{sec:conclusion}

We studied cold, dense $(1+1)$-dimensional $\mathrm{SU}(2)$ gauge theory with a single flavor.
Using uniform matrix product states, we constructed optimized ground states obtained by minimizing the grand potential at zero temperature with the VUMPS algorithm, maintaining gauge invariance.
From the ground states, we extracted the baryon number density, the pressure, and the quark distribution function.
Our results are consistent with previous studies, and the use of translationally invariant states eliminates undesirable boundary effects.

From the constructed ground states, we find the following properties.
We extract the baryon mass for zero baryon chemical potential.
The baryon number density develops a nonzero expectation value at a critical baryon chemical potential, $\mu_B=M_B$, where the threshold value $M_B$ depends on the quark mass $m$. 
We identify the transition point for fixed lattice input parameters (See Fig.~\ref{fig:thermodynamical quantities}).
A gapped phase is realized for $\mu_B<M_B$, while the system transitions to a gapless phase for $\mu_{B}>M_B$.
At zero baryon density with massless quarks, we provide a first-principles lattice determination of the central charge of $\mathrm{QC_2D_2}$, obtaining $c_\mathrm{IR} = 1.04$, in agreement with the conformal field theory prediction $c=1$.
At finite baryon density, the central charge is estimated to be $c=1$ within the fitting uncertainty.
This value coincides with that of the spinless Tomonaga--Luttinger liquid theory.

At finite density, the eigenvalues of the transfer matrix have nonzero complex phases. 
Consequently, the two-point function of the baryon number density exhibits the spatial modulation as a function of the relative distance. 
The corresponding wavenumber is consistent with the prediction of the Tomonaga--Luttinger liquid, $p=2\pi n_B\ell$ with $n_B$ and $\ell$ being the expectation value of baryon number density and the nonzero integer, respectively.
The speed of sound and the Luttinger parameter are evaluated from thermodynamic quantities such as the equation of state and the compressibility.
Our results show that the infrared behavior of dense $\mathrm{QC_2D_2}$ at finite baryon density is consistent with a Tomonaga--Luttinger liquid, and suggest the quarkyonic crossover scenario in $(1+1)$-dimensional non-Abelian gauge theory.

There are several interesting directions for future work.
Our simulation is performed for cold and dense $(1+1)$-dimensional two-color $\mathrm{QCD}$ with a single flavor, which provides a simplified and theoretically tractable setting to study QCD dynamics.
Although extending the analysis to higher spatial dimensions is a promising direction, it may still be challenging in several respects.
Another interesting direction, which is feasible with current computational resources, is to consider $(1+1)$-dimensional QCD with a general number of colors $N_c > 2$ and/or multiple flavors $N_f > 1$.
The infrared behavior of a single-flavor theory with a large number of colors $N_c > 2$ may be similar to that of the $N_c = 2$ case, since only the baryon-density mode remains in the long-distance limit, as suggested by analyses based on bosonization.
The multi-flavor extension may be even more interesting, in that non-Abelian global symmetries can persist in the infrared at finite density.
It would also be interesting to examine the quarkyonic picture more quantitatively by extending the present analysis to larger $N_c$~\cite{McLerran:2007qj,Kojo:2009ha}.
In particular, the Luttinger parameter is expected to approach $K \to 1/N_c$ in the quark-dominated regime, so that the separation between the baryonic gas regime, $K \simeq 1$, and the quarkyonic limit, $K \simeq 1/N_c$, becomes more pronounced as $N_c$ increases.

Another interesting direction is to investigate the quarkyonic shell structure~\cite{Kojo:2021wbc}, in which quark Pauli blocking depletes baryon occupation at low momenta.
Since baryons are bosonic diquarks for $N_c = 2$, a meaningful test requires extending the analysis to $N_c = 3$, where baryons are fermionic three-quark states and their momentum distribution can be constructed from the two-point function of a gauge-invariant baryon operator.

\vspace{4mm}
\acknowledgments
A part of the numerical simulations has been performed using Yukawa-21 at YITP in Kyoto University.
K.F. thanks Tomoya Hayata, Yuya Tanizaki, Takuya Okuda, and Yui Hayashi for useful comments and discussions.
This work was supported by JST, CREST Grant Number JPMJCR24I3.
K.F. was also supported by JSPS KAKENHI Grants No.~JP26K17144.
Y.H. was also supported by JSPS KAKENHI Grants No.~JP24H00975 and No.~JP25K01002.
The authors thank the RIKEN Center for Interdisciplinary Theoretical and Mathematical Sciences. Discussions during ``New computational methods in quantum field theory 2026'' were useful in completing this work.

\appendix

\section{A bosonic description of single-flavor \texorpdfstring{$\mathrm{QCD}_2$}{QCD2}}\label{appendix:bosonization}

There are no dynamical degrees of freedom of the gauge field in $(1+1)$-dimensions.
Consequently, the gauge field can be integrated out under a certain gauge fixing condition, and the resultant theory is purely described by the scalar field, as we shall see below.
Although our numerical simulations are performed in two-color QCD, we keep $N_c$ as a free parameter to make the color dependence explicit.
A peculiarity of two-color QCD is also noted.
Throughout this work, we assume a single flavor.

The action of the $\mathrm{SU}(N_c)$ gauge theory with a single fundamental fermion in two dimensions is given by 
\begin{align}
    S=\int\mathrm{d}^2x\left[-\frac{1}{2}\mathrm{Tr}\,(F^{\mu\nu}F_{\mu\nu})+i\bar{q}\slashed{D}q+\mu_qq^\dag q-m_q\bar{q}q\right].\label{eq:QCD action}
\end{align}
Here, we use the notation
\begin{align}
    \eta_{\mu\nu}=\mathrm{diag}(+1,-1),~\epsilon^{01}=-\epsilon_{01}=1,~\gamma^0=\sigma_1,~\gamma^1=i\sigma_2,~\gamma^5=\gamma^0\gamma^1=-\sigma_3.
\end{align}
$F_{\mu\nu}=\partial_\mu A_\nu-\partial_\nu A_\mu+ig_0[A_\mu,A_\nu]$ is the field strength of the gauge field $A_\mu=A_{\mu}^aT_a$, with $T_a$ and $g_0$ being the generators of the $\mathrm{SU}(N_c)$ group in the fundamental representation and gauge coupling, respectively.
The covariant derivative is given by $\slashed{D}=\gamma^\mu(\partial_\mu+ig_0A_{\mu})$.
The matrices $\sigma_i$ denote the Pauli matrices.
The gamma matrices satisfy the standard anti-commutation relation $\{\gamma^\mu,\gamma^\nu\}=2\eta^{\mu\nu}$, and the above choice is suitable for later purposes.
$\mu_q$ and $m_q$ are the quark chemical potential and the quark Dirac mass, respectively.

Let us note a peculiarity of flavor symmetries of two-color QCD.
Because the fundamental representation of $\mathrm{SU}(2)$ is pseudo-real, the flavor symmetry is enlarged.\footnote{See~\cite{Kogut:2000ek} for discussions of flavor symmetry in the case of $3+1$ dimensional two-color QCD.}
To see this, we first consider the kinetic term of quarks, given by
\begin{align}
    S_{\mathrm{kinetic}}&=\int \mathrm{d}^2x\left[i\bar{q}\slashed{D}q\right]=\int\mathrm{d}^2x\,i\left[q_L^\dag D_Lq_L+q_R^\dag D_Rq_R\right]\nonumber\\
    &=\dfrac{i}{2}\int\mathrm{d}^2x\left[q_L^\dag D_Lq_L-(D_Lq_L)^\dag q_L + q^\dag_RD_Rq_R -(D_Rq_R)^\dag q_R\right],
\end{align}
where $D_L=D_0-D_1$ and $D_R=D_0+D_1$.
In the second line, we have performed integration by parts on half of the terms.
Owing to the pseudo-reality of the $\mathrm{SU}(2)$ representation, it is convenient to introduce the following multiplet:
\begin{align}
    &q^\dag_{L(R)}D_{L(R)} - (D_{L(R)} q_{L(R)})^\dag q_{L(R)}=
    -\begin{pmatrix}
        q_{L(R)}& \tilde{q}_{L(R)}
    \end{pmatrix}
    \begin{pmatrix}
        0 & 1\\
        -1 & 0
    \end{pmatrix}
    \begin{pmatrix}
            \epsilon D_{L(R)} & 0\\
        0 &  \epsilon D_{L(R)}
    \end{pmatrix}
    \begin{pmatrix}
    q_{L(R)}\\
    \tilde{q}_{L(R)}
    \end{pmatrix}.
\end{align}
Here, we define $\tilde{q}_{L(R)\, c_1}=\epsilon_{c_1c_2}q^{\dag c_2}_{L(R)}$.
In this expression, we use $(D_{L(R)}\tilde{q}_{L(R)})_{c_1}=\epsilon_{c_1c_2}(D_{L(R)}q_{L(R)})^{\dag c_2}$, which follows from $\epsilon T_a \epsilon^{-1} = -(T_a)^T$, where $T_a$ are generators of $\mathrm{su}(2)$.
Therefore, there exists a flavor symmetry under 
\begin{align}
Q_{L(R)}\to U_{\mathrm{USp}(2)_{L(R)}}Q_{L(R)},
\end{align}
where $Q_{L(R)}=(q_{L(R)},\tilde{q}_{L(R)})^T$, and $U_{\mathrm{USp}(2)_{L(R)}}$ are elements of the unitary symplectic group.\footnote{A unitarity condition may be required to ensure a canonical anti-commutation relation \eqref{eq:canonical anti-commutation relation}.}
The $\mathrm{U}(1)_B$ group associated with the transformation $q_{L(R)}\to e^{i\theta_B/2}q_{L(R)}$ and $\tilde{q}_{L(R)}\to e^{-i\theta_B/2}\tilde{q}_{L(R)}$ is a subgroup of $\mathrm{USp}_L(2)\times \mathrm{USp}_R(2)$.
The center of the $\mathrm{SU}(2)$ gauge group acts as $q_{L(R)}\to -q_{L(R)}$ and $\tilde{q}_{L(R)}\to -\tilde{q}_{L(R)}$,
which causes redundancy of the flavor symmetry.
As a result, the flavor symmetry of single-flavor massless two-color QCD in $(1+1)$ dimensions is
$\mathrm{USp}(2)_L\times \mathrm{USp}(2)_R/\mathbb{Z}_2$.
The extension to $N_f$ flavors is straightforward, which leads to $\mathrm{USp}(2N_f)_L\times \mathrm{USp}(2N_f)_R/\mathbb{Z}_2$.

We next examine how this symmetry is affected by the mass term:
\begin{align}
    m_q \bar{q}q=
    m_q 
    \begin{pmatrix}
    q_R &\tilde{q}_R
    \end{pmatrix}
    \begin{pmatrix}
        0&1\\
        1&0
    \end{pmatrix}
    \epsilon
    \begin{pmatrix}
    q_L\\
    \tilde{q}_L
    \end{pmatrix}.
\end{align}
Under the $\mathrm{USp}(2)_L \times \mathrm{USp}(2)_R$, this term transforms as
\begin{align}
m_q
    \begin{pmatrix}
    q_R &\tilde{q}_R
    \end{pmatrix}
    \begin{pmatrix}
        0&1\\
        1&0
    \end{pmatrix}
    \epsilon
    \begin{pmatrix}
    q_L\\
    \tilde{q}_L
    \end{pmatrix}
    \to m_q
    \begin{pmatrix}
    q_R &\tilde{q}_R
    \end{pmatrix}
    U_{\mathrm{USp}(2)_R}^T
    \begin{pmatrix}
        0&1\\
        1&0
    \end{pmatrix}
    U_{\mathrm{USp}(2)_L}
    \epsilon\,
    \begin{pmatrix}
    q_L\\
    \tilde{q}_L
    \end{pmatrix}.
\end{align}
Therefore, the mass term breaks $\mathrm{USp}(2)_L \times \mathrm{USp}(2)_R / \mathbb{Z}_2$ down to the subgroup satisfying
\begin{align}
    U_{\mathrm{USp}(2)_R}^T \sigma_1 U_{\mathrm{USp}(2)_L} = \sigma_1 .
\end{align}
This condition implies
\begin{align}
U_{\mathrm{USp}(2)_R} = \sigma_3 U_{\mathrm{USp}(2)_L} \sigma_3,
\end{align}
and hence the unbroken subgroup is isomorphic to $\mathrm{USp}(2)$.
Note that $\mathrm{USp}(2) \simeq \mathrm{SU}(2)$.
Accordingly, the remaining flavor symmetry is $\mathrm{USp}(2)/\mathbb{Z}_2\simeq \mathrm{SU}(2)/\mathbb{Z}_2$.
Consequently, physical states decompose into irreducible representations of this group.

Similarly, we can examine the effect of a baryon chemical potential:
\begin{align}
    \mu_q q^\dag q=\mu_q\left(\tilde{q}_L\epsilon q_L+\tilde{q}_R \epsilon q_R\right),
\end{align}
which preserves $\mathrm{U}(1)_B/\mathbb{Z}_2$.

We now return to the discussion for a general number of colors, $N_c$.
The canonical commutation relations for the gauge fields are given by 
\begin{align}
    &[A_{0}^a(t,x_1),\Pi^{0}_b(t,x_2)]=i\delta^{a}_b\delta(x_1-x_2),\quad\Pi^{0}_a\equiv\dfrac{\delta S}{\delta (\partial_0A_{0}^a)} \approx 0,\\
    &[A_{1}^a(t,x_1),\Pi^{1}_b(t,x_2)]=i\delta^{a}_b\delta(x_1-x_2),\quad~\Pi^{1}_a(t,x)\equiv\dfrac{\delta S}{\delta (\partial_0A_{1}^a)}=-F^{01}_{a}.
\end{align}
Here, $a=1,2,\cdots,N_c^2-1$ is the color index in the adjoint representation, and $\approx$ implies that the relation holds after substituting the canonical quantization condition.
The quark anti-commutation relation is given by
\begin{align}
    \{q_{c_1}(t,x_1),\pi^{c_2}_{q}(t,x_2)\}=i\delta^{c_1}_{c_2}\delta(x_1-x_2), \label{eq:canonical anti-commutation relation}
\end{align}
where $\pi^{c_1}_q(t,x)\coloneqq iq^{\dag c_1}(t,x)$.
In this expression, $c=1,2,\cdots,N_c$ represents the color index in the fundamental representation.

The Hamiltonian at zero-temperature is constructed from the action eq.~\eqref{eq:QCD action} and is given by
\begin{align}
    H = \int{\mathrm d}x\left[\frac{1}{2}E_a^2-\bar{q}i\gamma^1D_1q+\bar{q}(-\gamma^0\mu_q+m_q)q-A^a_{0}((D_\mathrm{adj}E)_a-\rho_a)\right]. \label{eq:grand potential}
\end{align}
We note that this expression includes the chemical potential term; more precisely, it corresponds to $H - \mu_B N_B$ in the main text.
In this expression, we rewrite the electric field by $E_a= \Pi^1_{a}$.
$(D_\mathrm{adj}E)_a= \partial_1E_a-g_0A^b_{1}f_{ab}^{~~c}E_c$ and $\rho_{a}=g_0q^\dag T_aq$ are the covariant derivatives in the adjoint representation for the spatial direction and the color charge density, respectively.
Because $\Pi_{0a}\approx0$, time evolution of the operator $\Pi_{0a}$ should satisfy the following constraint,
\begin{align}
    \partial_0\Pi_{0a}=-i[H,\Pi_{0a}]\approx 0.
\end{align}
Hence, the physical state must obey the Gauss-law constraint,
\begin{align}
((D_\mathrm{adj}E)_a-\rho_a)|\psi\rangle=0,~(D_\mathrm{adj}E)_a= \partial_1E_a-g_0A^b_{1}f_{ab}^{~~c}E_c,~\rho_{a}=g_0q^\dag T_aq. \label{eq:Gauss-law constraint continuum}
\end{align}
In the following discussion, we do not consider the operator $A_{0a}$ except for the Hamiltonian, and thus, $\approx$ is replaced by the equality.

The Gauss-law constraint is simplified under the axial gauge condition $A_1=0$.
In this gauge, we have
\begin{align}
    \partial_1E_a-\rho_a= 0.
\end{align}
We put the system on a finite interval $x\in[-L,L]$ and take the boundary condition such that the electric field vanishes at the boundary, $E_a(x=\pm L)=0$.
Then the electric field is given by 
\begin{align}
    E_a(x)=\int^x_{-L}\mathrm{d}z\,\rho_a(z).
\end{align}
Note that
\begin{align}
    \int^L_{-L}\mathrm{d}z\,\rho_a(z)=0 \label{eq:total charge neutrality condition}
\end{align}
follows from the boundary condition, $E_a(L)=0$.
The square of the total electric field energy can be eliminated as follows:
\begin{align}
    \int^L_{-L} \mathrm{d}x\,E_a^2(x)&=\int^L_{-L} \mathrm{d}x\int^x_{-L}\mathrm{d}z\int^x_{-L}\mathrm{d}w\,\rho_a(z)\rho_a(w)\nonumber\\
    &=\int^L_{-L} \mathrm{d}x\int^{L}_{-L}\mathrm{d}z\int^L_{-L}\mathrm{d}w\,\Theta(x-z)\Theta(x-w)\rho_a(z)\rho_a(w)\nonumber\\
    &=\int^{L}_{-L}\mathrm{d}z\int^L_{-L}\mathrm{d}w\,\left(L-\frac{z+w+|z-w|}{2}\right)\rho_a(z)\rho_a(w)\nonumber\\
    &=-\frac{1}{2}\int^L_{-L}\mathrm{d}z\int^L_{-L}\mathrm{d}w\,|z-w|\rho_a(z)\rho_a(w).\label{eq:electric flux tube}
\end{align}
In the last equality, we use the total charge neutrality condition, eq.~\eqref{eq:total charge neutrality condition}.
One can see from the above expression, two color charge densities of matter fields experience linear confinement.
The resultant Hamiltonian is given by
\begin{align}
    H= \int^L_{-L}{\mathrm d}x\left[-\bar{q}i\gamma^1\partial_1q+\bar{q}(-\gamma^0\mu_q+m_q)q\right]-\frac{1}{4}\int^L_{-L}\mathrm{d}x_1\int^L_{-L}\mathrm{d}x_2\,|x_1-x_2|\rho_a(x_1)\rho_a(x_2).
\end{align}
Although the axial gauge condition is suitable for illustrating linear confinement, it is not well suited for the discussion based on the Abelian bosonization presented below.
Therefore, we adopt a different gauge-fixing condition.

We now apply bosonization.
While non-Abelian bosonization~\cite{Witten:1983ar,Frishman:1992mr} is generally required for multi-flavor QCD, Baluni proposed a gauge-fixing condition that allows one to study the single-flavor case using simple Abelian bosonization~\cite{Baluni:1980bw}.
Based on this approach, Steinhardt analyzed the mass spectrum of single-flavor $\mathrm{QCD}_2$ in the strong-coupling regime $g_0 \gg m_q$, assuming that all colored sectors in the bosonic description can be safely integrated out~\cite{Steinhardt:1980ry}.

In our case, there are $N_c$ Dirac fermions and correspondingly $N_c$ compact bosons.
They are related as
\begin{align}
    &q_c=\begin{pmatrix}
    L_c\\
    R_c
    \end{pmatrix},\nonumber\\
    &L_c=\dfrac{\eta_{L,c}}{\sqrt{2\pi a_\epsilon}}:\exp\left(-i\phi_{L,c}(t+x)\right):,\label{eq:bosonization formula}\\
    &R_c=\dfrac{\eta_{R,c}}{\sqrt{2\pi a_\epsilon}}:\exp\left(i\phi_{R,c}(t-x)\right):,\nonumber
\end{align}
where the scheme-dependent cutoff $a_\epsilon$ is introduced.
The color dependent Klein factors $\eta_{L(R),c}$ ensure anti-commutation relations~\eqref{eq:canonical anti-commutation relation}.
$:\cdots:$ denotes normal-ordering with respect to the compact boson with cutoff $a_\epsilon$.
$\phi_{L(R),c}$ are the chiral bosons, which correspond to the left (right) moving fields.
In our formulation, the periodicity is given by $\phi_{L(R)}\sim \phi_{L(R)}+2\pi$.

The Gauss-law constraint \eqref{eq:Gauss-law constraint continuum} can be rewritten as
\begin{align}
    &(\partial_1E)_{c_1}^{~c_2}+ig_0([A,E])_{c_1}^{~c_2}=\frac{g_0}{2}\left(q^{\dag c_2}q_{c_1} -\frac{1}{N_c}q^{\dag c} q_c\,\delta_{c_1}^{~c_2}\right),
\end{align}
where we expand a matrix-valued field as $X_{c_1}^{~c_2}=X_a(T^a)_{c_1}^{~c_2}$ with $X$ being the electric or gauge fields.
Using gauge transformations, $E\to U(x)EU^\dag(x)$ where $U(x)\in \mathrm{SU}(N_c)$, one can diagonalize the electric field.
Furthermore, there remain residual gauge transformations of the form $U(x)=\mathrm{diag}(e^{i\alpha_1(x)},e^{i\alpha_2(x)},\cdots,e^{i\alpha_{N_c}(x)})$ with $\sum_j\alpha_j(x)=0$.
These can be used to eliminate $N_c-1$ independent diagonal components of the gauge field.
Therefore, we can work in the following basis:
\begin{align}
    E_{c_1}^{~c_2}=e_{c_1}\delta_{c_1}^{~c_2},\quad(A_1)_{c}^{~c}=0.\label{eq:Baluni gauge condition}
\end{align}
In the second expression, we do not take the sum with respect to $c$.
Under this gauge-fixing condition, the Gauss-law constraint can be decomposed into diagonal and off-diagonal components as
\begin{align}
    \partial_1e_{c}&=\frac{g_0}{2}\left(j_{c}^0-j^0_B\right),\label{eq:diagonal electric field}\\
    -ig_0(e_{c_1}-e_{c_2})(A_1)_{c_1}^{~c_2}&=\frac{g_0}{2}(j^0)_{c_1}^{~c_2},
\end{align}
where 
\begin{align}
    j_{c}^\mu=\bar{q}^{c}\gamma^\mu q_{c},\quad
    j^{\mu}_B=\frac{1}{N_c}\sum_{c=1}^{N_c}\bar{q}^{c}\gamma^\mu q_c,\quad
    (j^\mu)_{c_1}^{~c_2}=\bar{q}^{c_2}\gamma^\mu q_{c_1}.
\end{align}
Using the bosonization, the diagonal components of the Gauss-law constraint lead to
\begin{align}
    e_{c} = \frac{g_0}{4\pi}(\phi_c-\phi_B),\quad\phi_B=\frac{1}{N_c}\sum_c\phi_c,
\end{align}
with $\phi_c\coloneqq\phi_{L,c}+\phi_{R,c}$, assuming there is no color density at boundaries.
Because of these boundary conditions, the periodicity of $\phi_c$ no longer holds.
The off-diagonal components of the Gauss-law constraint can be used to eliminate $A_1$.\footnote{One cannot eliminate $A_1$ when there exists degeneracy, $e_{c_i}=e_{c_j}$. In this case, we obtain constraint $(j^0)_{c_i}^{~c_j}=0$. This relation eliminates the Hamiltonian density involving $A_1$, and thus our discussion on the colored sector is unchanged.}

The Hamiltonian density of the electric field becomes
\begin{align}
    \mathcal{H}_\mathrm{ele}=\sum_{a=1}^{N_c^2-1}\frac{E_a^2}{2}= \sum_{c=1}^{N_c}e_{c}^2=\frac{1}{N_c}\sum^{N_c}_{c_1>c_2}(e_{c_1}-e_{c_2})^2=\frac{g_0^2}{16\pi^2 N_c}\sum_{c_1>c_2}^{N_c}\left(\phi_{c_1}-\phi_{c_2}\right)^2,
\end{align}
where we have used the constraint $\sum_ce_{c}=0$ in the third and fourth equalities.
Note that the Hamiltonian density of the electric field becomes a mass term for the colored sector, $\phi_{c_1}-\phi_{c_2}$.
Moreover, the Hamiltonian density involving the gauge field can be rewritten as
\begin{align}
    \mathcal{H}_{A_1}=g_0\bar{q}^{c_1}\gamma^1q_{c_2} (A_1)_{c_1}^{~c_2}=\sum_{c_1>c_2}^{N_c}\frac{ig_0}{2(e_{c_1}-e_{c_2})}((j^1)_{c_2}^{~c_1}(j^0)_{c_1}^{~c_2}-(j^1)_{c_1}^{~c_2}(j^0)_{c_2}^{~c_1})=\sum_{c_1>c_2}^{N_c}f(\phi_{c_1}-\phi_{c_2}).\label{eq:Hamiltonian density of gauge field}
\end{align}
Here, $f(\phi_{c_1}-\phi_{c_2})$ is expressed by $\sin(\phi_{c_1}-\phi_{c_2})$ with a scheme-dependent normalization.
The most important point is that the colored sectors such as the electric and the gauge fields depend only on the differences $\phi_{c_1}-\phi_{c_2}$ ($c_1\neq c_2$) once the gauge condition~\eqref{eq:Baluni gauge condition} is imposed.

Because the remaining terms in the Hamiltonian density are diagonal in the color indices, it is straightforward to perform the Abelian bosonization.
In particular, we employ the following formulas~\cite{Coleman:1974bu,Mandelstam:1975hb,Frishman:1992mr}:
\begin{align}
    :\bar{q}^cq_c:\,&=-\frac{1}{\pi a_\epsilon}\cos(\phi_c),\label{eq:bosonization mass}\\
    :\bar{q}^c\gamma^\mu q_c:\,&=\frac{1}{2\pi}\epsilon^{\mu\nu}\partial_\nu \phi_c,\label{eq:bosonization vector current}\\
    :-i\bar{q}^c\gamma^1\partial_1q_c:\,&=\frac{1}{8\pi}\left((\partial_0\phi_c)^2+(\partial_1\phi_c)^2\right).
    \label{eq:bosonization kinetic term}
\end{align}
Here, no summation over $c$ is implied.
From eq.~\eqref{eq:bosonization kinetic term}, the kinetic part of the Hamiltonian density is given by
\begin{align}
    \mathcal{H}_\mathrm{kin}&=\sum_{c=1}^{N_c}:-i\bar{q}^c\gamma^1\partial_1q_c:\,\notag\\
    &=
    \dfrac{1}{8\pi }\sum^{N_c}_{c=1}\left(
(\partial_0\phi_{c})^2+(\partial_1\phi_{c})^2
        \right)
        \notag\\
    &=\dfrac{1}{8\pi N_c}\sum^{N_c}_{c_1>c_2}\left((\partial_0\phi_{c_1,c_2})^2+(\partial_1\phi_{c_1,c_2})^2\right)+\dfrac{N_c}{8\pi}\left((\partial_0\phi_{B})^2+(\partial_1\phi_{B})^2\right),
\end{align}
where we introduce
\begin{align}
    &\phi_{c_1,c_2}=\phi_{c_1}-\phi_{c_2},\quad
    \phi_B = \frac{1}{N_c}\sum_{c=1}^{N_c}\phi_c.
\end{align}
Here, we explicitly separate the colored sector and the singlet baryon density sector.
The Hamiltonian densities involving the Dirac mass and the quark chemical potential are obtained from eqs.~\eqref{eq:bosonization mass} and \eqref{eq:bosonization vector current} as  
\begin{align}
    &\mathcal{H}_{m_q}=m_q\sum_{c=1}^{N_c}:\bar{q}^cq_c:= -\dfrac{m_q}{\pi a_{\epsilon}}\sum^{N_c}_{c=1} \cos\left(\phi_c\right),\label{eq:Hamiltonian density of the Dirac mass term}\\
    &\mathcal{H}_{\mu_B}=-\mu_q\sum_{c=1}^{N_c}:q^{\dag c} q_c: =-\frac{\mu_B}{2\pi}\partial_1 \phi_B,
\end{align} 
where we introduce the baryon chemical potential $\mu_B =N_c\mu_q$.
The total Hamiltonian density can be decomposed into the free part $\mathcal{H}_\mathrm{free}$ and interacting $\mathcal{H}_\mathrm{int}$ part in the bosonized description as 
\begin{align}
    &\mathcal{H}_\mathrm{tot}=\mathcal{H}_\mathrm{free}+\mathcal{H}_\mathrm{int},\\
    &\mathcal{H}_\mathrm{free}=\mathcal{H}_\mathrm{kin} +\mathcal{H}_\mathrm{ele} + \mathcal{H}_{\mu_B},\\
    &\mathcal{H}_\mathrm{int}=\mathcal{H}_{A_1}+ \mathcal{H}_{m_q}.
\end{align}

It is convenient to introduce $(N_c-1)$ compact scalar fields orthogonal to the baryon mode through the relation,
\begin{align}
\phi_{(k)}\coloneqq\frac{1}{\sqrt{k(k+1)}}\left(\sum_{c=1}^{k}\phi_c-k\,\phi_{k+1}\right),
\qquad k=1,2,\dots,N_c-1.
\end{align}
Conversely, $\phi_c$ can be expressed as
\begin{align}
    &\phi_c=\phi_B+\sum_{k=c}^{N_c-1}\frac{1}{\sqrt{k(k+1)}}\phi_{(k)}-\sqrt{\frac{c-1}{c}}\;\phi_{(c-1)},\quad(N_c>c>1).
\end{align}
The last term is absent for $c=1$, while the second term is absent for $c=N_c$.
Note that $\sum_{c=1}^{N_c} \phi^2_c=N_c\phi_B^2+\sum_{k=1}^{N_c-1}\phi^2_{(k)}$.
It follows from this definition that
\begin{align}
    \sum_{c_1>c_2}^{N_c}(\phi_{c_1}-\phi_{c_2})^2=N_c\sum^{N_c-1}_{k=1}\phi_{(k)}^2.
\end{align}
In the basis $(\phi_B,\phi_{(k)})$, the free Hamiltonian is expressed as
\begin{align}
    \mathcal{H}_\mathrm{free}= \frac{1}{2R_c^2}\Pi_B^2
    +\frac{R_c^2}{2}(\partial_1\phi_B)^2
    -\frac{\mu_B}{2\pi}\partial_1 \phi_B
    +\sum_{k=1}^{N_c-1}\left[\frac{1}{2R^2}\Pi_{(k)}^2+\frac{R^2}{2}(\partial_1\phi_{(k)})^2 + \frac{g_0^2}{16\pi^2}\phi_{(k)}^2\right],
\end{align}
where
\begin{align}
    \Pi_B=R^2_c\partial_0\phi_B,\quad\Pi_{(k)}=R^2\partial_0\phi_{(k)},\quad
    R_c=\sqrt{\dfrac{N_c}{4\pi}},\quad R=\sqrt{\dfrac{1}{4\pi}}.
\end{align}
Here, $R_c$ is the radius of the compact boson associated with the $\mathrm{SU}(N_c)$ singlet baryon density mode.
The remaining $(N_c-1)$ compact bosons associated with colored sector acquire a mass $g_0/\sqrt{2\pi}$ in the basis where the kinetic term is canonically normalized.

For generic values of $m_q,~\mu_q$ and $\mu_B$, $\mathrm{QCD}_2$ cannot be solved analytically even after the bosonization due to the presence of nonlinear interactions arising from \eqref{eq:Hamiltonian density of gauge field} and \eqref{eq:Hamiltonian density of the Dirac mass term}.
In the limit of $g_0\gg m_q$, one can integrate out the colored sector and obtain an effective theory for the gauge singlet boson $\phi_B$.
To this end, we consider normal ordering with respect to the bare mass scalar of the colored sector, $\mu_{g_0}=g_0/\sqrt{2\pi}$.
We apply the same definition of the normal-ordering in ref.~\cite{Coleman:1974bu} with an identification of the cutoff scale, $\Lambda \sim 1/{a_\epsilon}$.
We then perform the renormal-ordering and obtain
\begin{align}
    \mathcal{H}_{m_q}&= -\dfrac{m_q}{\pi a_{\epsilon}}\sum^{N_c}_{c=1} \cos\left(\phi_c\right)\simeq -\frac{N_c}{\pi}m_q\left(\tilde{\mu}_{g_0}\right)^{(N_c-1)/N_c}a_\epsilon^{-1/N_c}\cos\phi_B = -M^2N_M[\cos\phi_B],
\end{align}
where
\begin{align}
        \tilde{\mu}_{g_0}&=\frac{g_0}{\sqrt{2\pi}}\frac{e^{\gamma_E}}{2},~M^2=\left(\frac{N_c}{\pi}m_q\tilde{\mu}_{g_0}^{1-1/N_c}\right)^{2N_c/(2N_c-1)}\left(\frac{e^{\gamma_E}}{2}\right)^{2/(2N_c-1)}.
\end{align}
In the second step, we approximate $N_{\mu}[\cos(\phi_{k})]= 1$, which is justified for $g_0\gg m_q$.
We can take an arbitrary mass scale $\mu_2$ for the normal-ordering using the following  relation,
\begin{align}
    N_{\mu_1}[e^{i\phi}]=\left(\dfrac{\mu_2}{\mu_1}\right)^{1/(4\pi \beta^2)}N_{\mu_2}[e^{i\phi}],\label{eq:scaling of normal-ordering}
\end{align}
where $\beta$ is the radius of the compact boson.
The remaining term $\mathcal{H}_{A_1}$ is neglected under the assumption that the Fock vacuum of the colored sector provides a good approximation.

Under this simplification, we arrive at the following expression,
\begin{align}
    H_\mathrm{boson}= \int\mathrm{d}x\,\left[\frac{1}{2R_c^2}\Pi_B^2+\frac{R_c^2}{2}(\partial_1\phi_B)^2-\frac{\mu_B}{2\pi}\partial_1\phi_B+M^2\left(1-N_M[\cos\left(\phi_B\right)]\right)\right]. \label{eq:bosonized Hamiltonian}
\end{align}
In this normalization, the baryon current is given by
\begin{align}
    j_{B}^\mu=\frac{1}{N_c}\sum_{c=1}^{N_c}:\bar{q}^{c}\gamma^\mu q_{c}:\,= \frac{1}{2\pi}\epsilon^{\mu\nu}\partial_\nu \phi_B,\label{eq:baryon number density bosonization}
\end{align}
and satisfies the canonical commutation relation,
\begin{align}
    [\phi_B(t,x),\Pi_B(t,y)]=i\delta(x-y).
\end{align}

For $\mu_B=0$, the above Hamiltonian reduces to the sine-Gordon model.
The lowest excitation is kink (and anti-kink), which carries the baryon (and anti-baryon) number $+1$ (and $-1$),
\begin{align}
    N_B=\int^\infty_{-\infty}\mathrm{d}x\,j_{B}^0=\frac{1}{2\pi}\int^\infty_{-\infty}\mathrm{d}x\, \partial_1\phi_B=\pm 1.
\end{align}
Higher excitations are argued to be bound states of baryon and anti-baryon~\cite{Steinhardt:1980ry}.
For $m_q=0$ and $\mu_q =0$, the theory reduces to a compact free scalar and becomes a conformal field theory with central charge $c=1$.

Let us turn to the case with a nonzero baryon chemical potential, $\mu_B\neq 0$.
It is useful to consider the extreme limit, $\mu_B \gg m_q$.
At leading order, one obtains
\begin{align}
    H = \int\mathrm{d}x\left[\frac{1}{2R_c^2}\Pi_B^2+\frac{R_c^2}{2}\left(\partial_1\phi_B -\dfrac{2\mu_B}{N_c}\right)^2-\dfrac{\mu_B^2}{2\pi N_c}\right].
\end{align}
The Hamiltonian is Gaussian and describes a single free compact boson.
The configuration $\langle \partial_1\phi_B\rangle_{m_q=0}=2\mu_B/N_c$ is favored to minimize the energy.
In terms of the baryon number density, one obtains
\begin{align}
    n_B= \langle j_B^0\rangle_{m_q=0} =\dfrac{\mu_B}{\pi N_c},
\end{align}
where we use the relation \eqref{eq:baryon number density bosonization}.
The pressure is obtained by integrating the baryon density with respect to the chemical potential:
\begin{align}
    P(\mu_B,m_q=0) =\dfrac{\mu_B^2}{2\pi N_c},
\end{align}
which coincides with $-H_{\Pi_B=0,\partial_1\phi_B=2\mu_B/N_c}$.

For $m_q\neq 0$ and $\mu_B\neq 0$, where the Hamiltonian is given by eq.~\eqref{eq:bosonized Hamiltonian}, it is not straightforward to reveal the infrared behavior, including the mass spectrum.
One can make a shift, $\phi_B=\tilde{\phi}_B+\langle\partial_1\phi_B\rangle x$ such that $\langle\tilde{\phi}_B\rangle = 0$.
In this basis, the system is effectively described by the Gaussian Hamiltonian with a cosine term, $\cos(\tilde{\phi}_B+\langle \partial_1\phi_B\rangle x)$.
When $\langle \partial_1\phi_B\rangle =0$, there is no modulation, and the cosine term generates a mass gap.
However, at finite density with $\langle\partial_1\phi_B\rangle\neq 0$, the cosine term can become irrelevant in the infrared due to the spatial modulation.
In this case, the system is expected to enter a gapless phase.
In fact, studies based on the thermodynamic Bethe ansatz indicate that a gapped phase is realized at low $\mu_B$, while a gapless phase emerges above a certain critical value of $\mu_B$.

Although the Hamiltonian is Gaussian in the gapless phase, the presence of the spatial modulation breaks Lorentz symmetry.
Thus, the effective Hamiltonian in the long-distance limit is expected to take the form
\begin{align} 
    H_\mathrm{TLL} (\mu_B>M_B) =\int\mathrm{d}x\,\left[2\pi vK \Pi_{B}^2+\dfrac{v}{8\pi K}(\partial_1\tilde{\phi}_B)^2\right],\label{eq:TLL Hamiltonian}
\end{align}
where we omit the contribution from the irrelevant operators.
The Luttinger parameter can be written in terms of the compactification radius as $R^2=1/(4\pi K)$.
The parameters $v(\mu_B,m_q)$ and $K(\mu_B,m_q)$ are the speed of sound and the Luttinger parameter, respectively, and should be determined by the matching condition.
In the present case, they receive the renormalization effect from $m_q$ and $\mu_B$.
The baryon number density operator can be expanded in terms of $\tilde{\phi}_B$ as
\begin{align}
    j_B^0=\left( n_B +\frac{\partial_1\tilde{\phi}_B}{2\pi}\right)\sum_{m=-\infty}^\infty\alpha_m :e^{im(\tilde{\phi}_B+2\pi n_B x)}:. \label{eq:baryon number density operator}
\end{align}
Here, normal-ordering is defined with respect to the Hamiltonian~\eqref{eq:TLL Hamiltonian}.
The coefficients $\alpha_m$ $(\alpha_{m=0}=1,~\alpha_m=\alpha_{-m})$ are the numerical constants which generically depend on short-distance effects and the regularization scheme.
We have used the fact that the baryon number density appears in the combination $\tilde{\phi}_B + 2\pi  n_B x$, and vertex operators $e^{im(\tilde{\phi}_B+2\pi n_B x)}$ are introduced such that correlation functions of theories defined by Hamiltonians~\eqref{eq:bosonized Hamiltonian} and \eqref{eq:TLL Hamiltonian} coincide.
 
Our heuristic derivation of the number density operator~\eqref{eq:baryon number density operator} strongly relies on the form of the microscopic description~\eqref{eq:bosonized Hamiltonian}.
Interestingly, this representation was found by Haldane in a universal manner~\cite{Haldane:1981zz,Haldane:1981zza}.
We now study the effective description at finite density, $ n_B\neq 0$ with an appropriate identification of a collective mode.
To keep the discussion transparent, we also give a brief derivation as follows.
We introduce a phenomenological baryon number density as
\begin{align}
    &j^0_B(t,x) = \sum_j\delta(x-x_j(t)).
\end{align}
Here, $j\in \mathbb{Z}$ labels the baryonic particles, and $x_j(t)$ denotes their positions.
The main goal is to rewrite the above baryon density operator in terms of a collective mode $\Phi_B(t,x)$.
We regard $\Phi_B(t,x)$ as a {\it slowly-varying} baryon number, which satisfies $\Phi_B(t,x)=j$ for $x=x_j(t)$.
One can then perform the following manipulation:
\begin{align}
    j^0_B(t,x)=\sum_j\delta(x-x_j(t))\simeq \partial_1\Phi_B(t,x)\sum_j\delta(\Phi_B(t,x)-j) =(\partial_1\Phi_B)\sum_{m=-\infty}^\infty e^{i2\pi m\Phi_B(t,x)}.
\end{align}
In the second step, we have assumed that $\Phi_B(t,x)$ is a smooth function.
The last equality follows from the standard Poisson summation formula.
By performing the shift $\Phi_B=\tilde{\Phi}_B+n_B x$ and taking normal ordering with respect to the free theory of $\tilde{\Phi}_B$, one can recover the density representation~\eqref{eq:baryon number density operator} up to the overall normalization.
We refer to ref.~\cite{Haldane:1981zz} for further discussions.
One of the great lessons from this derivation is that the sum over $m$ originates from the discreteness of the particle number labeled by $j$.

Once the representation of the baryon number density is obtained, we can evaluate the two-point function using the free Hamiltonian~\eqref{eq:TLL Hamiltonian}.
The result is given by 
\begin{align}
    \langle j^0_B(t,x_1)j^0_B(t,x_2)\rangle- n_B^2=\dfrac{B_0}{|x_1-x_2|^2} +\sum_{\ell=1}^{\infty} B_\ell\dfrac{\cos(2\pi \ell n_B |x_1-x_2|)}{|x_1-x_2|^{2\ell^2 K}}.
\end{align}
Here, $\langle \cdots \rangle$ denotes the ground-state expectation value with respect to the Hamiltonian~\eqref{eq:TLL Hamiltonian}.

An infinitesimal change in the baryon number density $n_B\to  n_B+\delta n_B$ is equivalent to the shift $\tilde{\phi}_B\to \tilde{\phi}_B+2\pi\delta n_Bx$, according to the expression of the baryon number operator, eq.~\eqref{eq:baryon number density operator}.
Hence, we obtain
\begin{align}
    \dfrac{\mathrm{d}^2\langle H_\mathrm{TLL} \rangle}{\mathrm{d}n_B^2}=\pi V\frac{v}{K},
    \label{eq:Luttinger1}
\end{align}
where $V$ denotes the spatial volume.
Here, $\langle H_\mathrm{TLL}\rangle$ should be interpreted as the total energy, because the expectation value of $\mu_BN_B$ is subtracted (whereas $\langle H_\mathrm{boson}\rangle$ corresponds to the grand potential).
Thus, the above relation can be rewritten as
\begin{align}
    \frac{\mathrm{d}\mu_B}{\mathrm{d}n_B}=\pi\frac{v}{K}.
    \label{eq:Luttinger2}
\end{align}
where we use the thermodynamic relation $\mathrm{d}\ed/\mathrm{d}n_B=\mu_B$, where $\ed$ is an energy density, $\langle H_\mathrm{TLL}\rangle/V$.
We assume that the speed of sound $v$ in the Hamiltonian eq.~\eqref{eq:TLL Hamiltonian} coincides with that defined from the equation of state:
\begin{align}
    c_s^2= \dfrac{\mathrm{d}P}{\mathrm{d}\ed}=\frac{\mathrm{d}P}{\mathrm{d}\mu_B}\frac{\mathrm{d}\mu_B}{\mathrm{d}\ed}=\frac{n_B}{\mu_B}\frac{\mathrm{d}\mu_B}{\mathrm{d}n_B}.
    \label{eq:Luttinger3}
\end{align}
Under this assumption, one can extract the speed of sound and the Luttinger parameter as
\begin{align}
   K= \frac{\pi}{c_s} \frac{n_B}{\mu_B}.
    \label{eq:Luttinger4}
\end{align}
For $m_q=0$, exact results are given by $n_B  =\mu_B/(\pi N_c),~v=1$ and $K=1/N_c$.
These results provide a useful consistency check for our numerical simulations.

Before closing this appendix, we discuss the behavior near the onset $n_B\to 0$, based on the effective Hamiltonian~\eqref{eq:bosonized Hamiltonian}.
It is known that this model (a Sine-Gordon model with a chemical potential, or equivalently to the massive Thirring model with a chemical potential) exhibits the Pokrovsky--Talapov transition at $\mu_B=M_B$, where $M_B$ is the mass of the kink excitation by classical analysis~\cite{1978JETP...48..579P,PhysRevLett.42.65}.
An essential observation is that kinks are equivalent to fermions so that the interaction among them is repulsive.
Hence, the thermodynamic behavior in the very low-density region may be described by the ideal gas of baryons (solitons)~\cite{1980JETP...51..134P}.
In this regime, the ground state is occupied by nonrelativistic kinks with momenta below $k_F=\pi  n_B$, which can be effectively described by a free fermion gas.
Within this approximation, the pressure is given by
\begin{align}
   P= -\int_{-k_F}^{k_F} \frac{\mathrm{d}k}{2\pi}\left[\frac{k^2}{2M_B}-\delta\mu_B\right]=-\dfrac{\pi^2n_B^3}{6M_B}+\delta\mu_Bn_B.
\end{align}
Here, $\delta\mu_B=\mu_B-M_B$.
The chemical potential is related to the Fermi momentum by $\mu_B=M_B+k_F^2/(2M_B)$.
Substituting this relation, the pressure and the energy density become
\begin{align}
    P=\frac{(2M_B\delta{\mu}_B)^{3/2}}{3M_B\pi}=\frac{\pi^2n_{B}^3}{3M_B},\quad
    \ed = M_Bn_B+\frac{\pi^2n_B^3}{6M_B}.
\end{align}
This expression implies that the second derivative of the pressure diverges at $\delta\mu_B=0$ as
\begin{align}
\frac{\mathrm{d}^2 P}{\mathrm{d} \mu_B^2} =\frac{\mathrm{d} n_B}{\mathrm{d}\mu_B}=\frac{\sqrt{M_B}}{\pi\sqrt{2\delta \mu_B}}
=\frac{M_B}{\pi^2n_B}.
\label{eq:EOS_near_MB}
\end{align}
This equation leads to the following sound velocity and Luttinger parameter:
\begin{align}
c^2_s&\simeq\frac{\pi^2 n_B^2}{M_B^2},\quad K= 1.
\end{align}
These expressions indicate that $c_s\to 0$ and $K\to 1$, as $n_B\to 0$.
Note that $K=1$ corresponds to the compactification radius of the compact boson equivalent to a free fermion.
When $n_B$ becomes larger and larger, neither the non-relativistic approximation nor the ideal-gas treatment can be justified.
According to this argument, the transition associated with the onset of baryon number density is a second-order phase transition.

The analysis based on the Bethe ansatz technique supports the above result and reveals the thermodynamics for an arbitrary baryon number density~\cite{Iwabuchi1995CIC,2001PhRvB..63h5109P}.
The subleading corrections to the free fermion gas approximation are also investigated in ref.~\cite{2001PhRvB..63h5109P}.
We confirm that the overall behavior of the baryon number density in our simulation is consistent with these results.
Unfortunately, the numerical data in the very low-density region is not sufficient to reveal the critical behavior near the transition point as we reported in the main text.

Almost all of the analyses based on the bosonization assume the effective Hamiltonian~\eqref{eq:bosonized Hamiltonian}, where the colored sector is integrated out.
In the strong coupling limit, this result provides a good approximation.
However, our simulation based on the Hamiltonian lattice gauge theory with the Gauss-law constraint is not restricted to this limit, and thus the colored sector is not necessarily decoupled.

\section{Definitions of local operators on the lattice}\label{appendix:local operator on lattice}

The lattice Hamiltonian is constructed in the following basis:
\begin{align}
    \gamma^{0}_{l}=\sigma_3,~\gamma^1_l=i\sigma_2,~\gamma_{l5}=\gamma^0_l\gamma^1_l=\sigma_1,~C=\sigma_2.
\end{align}
Here, $C$ is the charge conjugation matrix satisfying
\begin{align}
    C^{-1}\gamma_l^\mu C=-(\gamma_l^\mu)^T,
\end{align}
where $T$ represents the transpose.
From this relation, one can verify that $C^{-1}SC=(S^{-1})^T$, where $S=\exp(-i\omega_{\mu\nu}\Sigma^{\mu\nu}_l/4)$ ($\Sigma_l^{\mu\nu}=i[\gamma^\mu_l,\gamma^\nu_l]/2$) is a Lorentz transformation acting on the spinor field as $\psi\to S\psi$.
The quark (Dirac fermion) transforms under the parity as $q_c(t,x)\to \gamma^0_lq_c(t,-x)$.
Then, the operator
\begin{align}
    \Delta_B(t,x)= i\epsilon^{c_1 c_2}q^T_{c_1}C\gamma_{l5}q_{c_2},
\end{align}
is gauge invariant, and behaves as the parity-even scalar field having the unit baryon number.
Note that $(C\gamma_{l5})^T=C\gamma_{l5}$ in $(1+1)$ dimensions.\footnote{In contrast to this, $(C\gamma_5)^T=-(C\gamma_5)$ in four spacetime dimensions.
In this case, the above operator is identically zero for a single flavor two-color QCD.}
A lattice version of this baryon and its conjugate operators are used to detect the baryon and anti-baryon excitations from the ground state in our analysis.
The expectation value of $\Delta_B$ vanishes (even at finite baryon density) because spontaneous breaking of continuous symmetry is forbidden in $(1+1)$-dimensions.

The staggered fermion is implemented through the relation,
\begin{align}
    \psi_c(2n,2n+1)=\frac{i^{2n}}{\sqrt{2a}}
    \begin{pmatrix}
        \phi_c(2n)\\
        -i\phi_c (2n+1)
    \end{pmatrix}.
\end{align}
Thus, the lattice version of the diquark operator is given by
\begin{align}
    \Delta_B(2n,2n+1)= \frac{\epsilon^{c_1 c_2}}{2a}\left(\phi_{c_1}(2n)\phi_{c_2}(2n)+\phi_{c_1}(2n+1)\phi_{c_2}(2n+1)\right).
\end{align}
This diquark operator is used to detect baryonic excitations.

\bibliography{ref}

\end{document}